\documentclass[12pt]{article}

\usepackage{amsmath}
\usepackage{amsthm}
\usepackage{amsfonts}          
\usepackage{amssymb}           
\usepackage[all]{xy}           
\usepackage{epsfig}

\newlength{\xtrawidth}
\newlength{\xtraheight}
\newcommand{\eref}[1]{\eqref{#1}}
\newcommand{\sref}[1]{\S\ref{#1}}

\newcommand{\fref}[1]{Fig.~\ref{#1}}
\newcommand{\cref}[1]{Chapter~\ref{#1}}
\newcommand{\beq}{\begin{equation}}
\newcommand{\eeq}{\end{equation}}
\newcommand{\bea}{\begin{eqnarray}}
\newcommand{\eea}{\end{eqnarray}}
\newcommand{\bean}{\begin{eqnarray*}}
\newcommand{\eean}{\end{eqnarray*}}
\newcommand{\ba}{\begin{array}}
\newcommand{\ea}{\end{array}}

\def\IC{\mathbb{C}}
\def\II{\mathbb{I}}

\def\IR{\mathbb{R}}

\def\IZ{\mathbb{Z}}

\def\IP{\mathbb{P}}

\def\dim{{\rm dim}}
\def\cN{{\mathcal N}}
\def\cM{{\mathcal M}}

\def\cO{{\mathcal O}}

\newcommand{\tr}{\mathop{\mbox{Tr}}}
\newcommand{\res}{\mathop{\mbox{Res}}}
\newcommand{\sym}{\mbox{Sym}}
\newcommand{\hilb}{\ensuremath{\mbox{Hilb}}}
\newcommand{\odd}{\ensuremath{\mbox{odd}}}
\newcommand{\even}{\ensuremath{\mbox{even}}}
\newcommand{\floor}{\ensuremath{\mbox{floor}}}
\newcommand{\W}[1]{\widetilde{#1}}

\newcommand{\smat}[1]{{\scriptsize 
    \left( \begin{matrix}#1\end{matrix} \right)}}
\newcommand{\tmat}[1]{{\tiny \left( \begin{matrix}#1\end{matrix} \right)}}

\def\nn{\nonumber}

\newtheorem{theorem}{\bf THEOREM}

\newcommand{\setequation}{\setcounter{equation}{0}}

\newcommand{\cov}[1]{\mbox{Cover}( #1 )}
\newcommand{\gen}[1]{ \langle #1 \rangle}
\newcommand{\comment}[1]{}
\newcommand{\think}[1]{{\bf \sf \large ??? #1 ???}}

\begin{document}

\vskip 0.5in
%
\centerline{{\Huge Counting Gauge Invariants: the Plethystic Program}}
~\\

\renewcommand{\thefootnote}{\fnsymbol{footnote}}

\centerline{{\bf 
Bo Feng${}^{1,2}$\footnote{\tt b.feng@imperial.ac.uk} 
Amihay Hanany${}^{3}$\footnote{\tt ahanany@perimeterinstitute.ca}, 
Yang-Hui He${}^{4,5}$\footnote{\tt hey@maths.ox.ac.uk}, }}
{\small
\begin{flushleft}
${}^1${\it Blackett Lab.~\&
Inst.~for Math.~Science, Imperial College, London, SW7 2AZ, U.K.}\\
${}^2${\it Center of Mathematical Science, Zhejiang
University, Hangzhou 310027, P.~R.~China}\\
${}^3${\it
Perimeter Institute, 31 Caroline St. N., Waterloo, ON, N2L 2Y5, Canada
}\\
${}^4${\it 
Collegii Mertonensis in Academia Oxoniensis,
Oxford, 
OX1 4JD, U.K.} \\
${}^5${\it Mathematical
Institute, Oxford University, 24-29 St.\ Giles', Oxford, OX1 3LB,
U.K.}
\end{flushleft}
}

\setcounter{footnote}{0}
\renewcommand{\thefootnote}{\arabic{footnote}}
\vskip 0.8in

\begin{abstract}
We propose a programme for systematically counting the single and
multi-trace gauge invariant operators of a gauge theory. Key to this
is the plethystic function. We expound in detail the power of this
plethystic programme for 
world-volume quiver gauge theories of D-branes probing Calabi-Yau
singularities, an illustrative case to which the programme is not
limited, though in which a full intimate web of relations between the
geometry and the gauge theory manifests herself. We can also use
generalisations of Hardy-Ramanujan to compute the entropy of gauge
theories from the plethystic exponential. In due course, we also touch
upon fascinating connections to Young Tableaux, Hilbert schemes and
the MacMahon Conjecture.
\end{abstract}

\newpage

\tableofcontents

\vspace{1in}

\section{Introduction and Recapitulation}
Given a supersymmetric quantum field theory, one of the first
quantities one wishes to determine is the spectrum of BPS operators.
Such a desire becomes particularly manifest for the class of
theories which arise in the AdS/CFT correspondence in string theory.
Of special interest are chiral BPS mesonic operators of the
4-dimensional, $\cN=1$ SUSY gauge theory living on D3-branes probing
a Calabi-Yau (CY) singularity. Such a setup has been archetypal in
the aforementioned AdS/CFT correspondence and when the transverse CY
space is trivially $\IC^3$, we are in the paradigmatic $\cN=4$ CFT
and $AdS_5 \times S^5$ situation of \cite{Maldacena:1997re}. When
the transverse CY is non-trivial, we have new classes of so-called
quiver gauge theories, pioneered by \cite{DM}, which has been
extensively developed over the past decade (for a review,
q.v.~e.g.~\cite{rev}).

Of vital geometrical significance is the the fact that the BPS
mesonic operators form a chiral ring whose relations determine the
transverse Calabi-Yau geometry. More technically, the syzygy amongst
these gauge invariant operators (GIO's) (modulo F-flatness)
gives the equation of the
Calabi-Yau threefold as an affine variety. This correspondence is
guaranteed by the fact, {\it per construtio}, the D3-brane probe is
a point in the transverse CY. Thus an intimate relation is
established between the gauge theory and the algebraic geometry of
the transverse space.

In our recent paper \cite{Benvenuti:2006qr}, we solved the problem
of counting these mesonic GIO's for arbitrary singularities, both
single-trace and multi-trace, and for both large and finite number
of D3-branes. Using results from combinatorics, commutative algebra
and number theory, we advocate a {\bf plethystic programme} wherein
such counting problem is not only systematically addressed, but also
intrinsically linked to the underlying geometry. With a brief
recapitulation of this over-arching programme let us first occupy
the reader.

To set notation, let a stack of $N$ parallel coincident D3-branes
probe a Calabi-Yau singularity $\cM$. The mesonic BPS gauge invariant
operators fall into two categories: single- and multi-trace. The
former consists of words in operators, with gauge-indices contracted
but only a single overall trace and the latter, various products of
the single-trace GIO's. We let the generating function of the
single-trace GIO's be $f_N(t;~\cM)$, and that of the multi-trace be
$g_N(t;~\cM)$. The $n$-th coefficient in the power expansion for $f$
and $g$ would then give the number of GIO's at level $n$ (where level
can be construed as some representative $U(1)$ charge such as the
R-charge, in the problem. For simple cases like $\IC^n$ or the
conifold, a good $U(1)$ charge is the number of operators, but
generically it is not a good qauntum number and we will refer to a
typical U(1) charge). When there are enough isometries, such as in the
case of $\cM$ being a toric variety, we can refine the counting and
extend $f$ and $g$ to $f_N(t_1,t_2,t_3;~\cM)$ and
$g_N(t_1,t_2,t_3;~\cM)$. Power expansion in the variables $t_{1,2,3}$
again gives the number of GIO's, with the multi-degree now related to
global $U(1)$ charges of the problem, including R-charge and other
flavour charges. 
Some of the main results of \cite{Benvenuti:2006qr} are then as
follows.
\begin{itemize}

\item
The generating functions obey (we can easily generalise from a
single variable $t$ to the tuple $t_{i=1,2,3,\ldots}$):
\[
\fbox{$ g_1(t) = f_{\infty}(t); \qquad f_{\infty}(t) = PE[f_1(t)],
\quad g_{\infty}(t) = PE[g_1(t)]; \qquad g_{N}(t) = PE[f_N(t)] $}
\]
where $PE$ is the {\bf plethystic exponential} function defined as
\[ f(t) = \sum\limits_{n=0}^\infty a_n t^n \quad \Rightarrow \quad
g(t) = PE[f(t)] = \exp\left( \sum_{n=1}^\infty \frac{f(t^n) -
  f(0)}{n} \right) =
\frac{1}{\prod\limits_{n=1}^\infty (1-t^n)^{a_n}} \ .
\]

\item
The quantity $f_\infty = g_1$ is the geometric {\it point d'appui}
and can be directly computed from properties of $\cM$. We have
called it the (Hilbert-)Poincar\'e series. In
\cite{Benvenuti:2006qr}, we referred to this as the Poincar\'e
series; it is, in fact, more appropriate, for reasons which shall
become clear in \sref{s:hilbert}, to call it the {\bf Hilbert
series}, an appelation to which we henceforth adhere. 
When $\cM$ is
an orbifold $\IC^3 / G$ for some finite group $G$, $f_\infty$ is the
Molien series \cite{yauyu} (We remark that
Molien series and plethysms have appeared in the context of
four-dimensional dualities in \cite{Pouliot:1998yv}). 
When $\cM$ is a toric variety,
$f_\infty$ can be obtained from the toric diagram
\cite{Martelli:2006yb} (see also related
\cite{Martelli:2006vh,Butti:2006nk,Basu:2006id}). When $\cM$ is a
manifold of complete intersection $f_{\infty}$ can be directly
computed by the defining equations of the manifold.

\item
The inverse function to $PE$ is the plethystic logarithm, given by
\[ 
\hspace{-1in} f(t) = PE^{-1}(g(t)) = \sum_{k=1}^\infty
\frac{\mu(k)}{k} \log (g(t^k)) \ , \qquad \mu(k) := \left\{\ba{lcl}
0 & & k \mbox{ has repeated prime factors}\\
1 & & k = 1\\
(-1)^n & & k \mbox{ is a product of $n$ distinct primes} \ea\right.
\] 
where $\mu(k)$ is the M\"obius function. The plethystic logarithm
of the Hilbert series gives the syzygies of $\cM$, i.e., 
\[
\fbox{$f_1(t) = PE^{-1}[f_\infty(t)] = $\mbox{ defining equation of
$\cM$}.} 
\] 
In particular, if $\cM$ were complete-intersection,
$f_1(t)$ is a polynomial.

\item
For finite $N$, define the function $g(\nu ; t)$ such that
\[
f_\infty(t) = \sum\limits_{n=0}^\infty a_n t^n \quad \Rightarrow
\quad \fbox{$ g(\nu ; t) := \prod\limits_{n=0}^{\infty} \frac{1}{(1
- \nu  \, t^n)^{a_n}} = \sum\limits_{N=0}^\infty g_N(t) \nu^N $} \ .
\] 
In other words, the $\nu$-expansion of $g(\nu ; t)$ gives the
generating function $g_N(t)$ of multi-trace GIO's for finite number
$N$ of D3-branes. The single-trace generating function $f_N(t)$ is
then retrieved from $g_N(t)$ by $PE^{-1}$. This qualifies $\nu$ as
the chemical potential for the number of D3-branes.

Crucial to the derivation of the above expression is the almost
tautological yet very important fact that 
\[ g_N(t; \cM) = g_1(t;\sym^N(\cM)), \qquad \sym^N(\cM) := \cM^N/S_N \
. 
\] 
That is to say,
the moduli space of a stack of $N$ D3-branes is the $N$-th
symmetrised product of that of a single D3-brane, viz., the
Calabi-Yau space $\cM$. 
\end{itemize} 

The above points highlight the
key constituents of the plethystic programme and inter-relates the
D-brane quiver gauge theory and the geometry of $\cM$. Indeed, one
function distinguishes herself, viz., $f_\infty$, which, as a
Hilbert series, can be obtained directly from the geometry.
Henceforth, as was in \cite{Benvenuti:2006qr}, we will often denote
the fundamental generating function $f_\infty$ and its associated
$g_\infty$ simply as $f$ and $g$.

We emphasise that the applicability of the plethystic programme is
not limited to world-volume theories of D-brane probes on Calabi-Yau
singularities. Indeed, if we knew the geometry of the classical
moduli space of a gauge theory, which may not even be $\cN=1$, and
especially if this vacuum space is a complete intersection variety,
we could obtain the Hilbert series and thenceforth use the
plethystic exponential to find the gauge invariants.

Without much further ado, let us outline the contents of our current
paper. In \sref{s:explicit} we derive explicit expressions for the
plethystic exponential. We will see how to recursively write all
$g_N$ generating functions in terms of the fundamental Hilbert
series; natural connexions with Young tableaux will arise. Of great
importantance will also be the asymptotic behaviour of the
multi-trace generating functions $g_N$ and we will see how a result
due to Haselgrove and Temperley may be used to generalise the
Meinardus theorem. Thus armed, we can estimate the entropy of our
gauge theory; this is the subject of \sref{s:entropy}. We will
explicitly see the dependence of the critical exponents on the
dimension of the geometry and the volume of the Sasaki-Einstein
manifold.

With all this technology, we move on to concrete classes of
examples. In \sref{s:su2}, we analytically compute the number of
single-trace operators for the ADE-singularities and give the
expressions for the asymptotic behaviour of the number of
multi-trace operators. As a passing curiosity, we point out intimate
relations to the MacMahon Conjecture. Then, in \sref{s:su3}, we
compute all fundamental generating functions for Calabi-Yau
threefold orbifolds, again, in explicit detail. Subsequently, one
can allow discrete torsion in these cases, and see how the
plethystic programme also encompasses these classes of theories in
\sref{s:dis}. As a mathematical aside, we see how the
plethystics relate to Hilbert schemes of points in \sref{s:hilb}.
Finally, moving onto toric varieties, we see how the plethystic
programme lends itself to deriving the equations for wide classes of
moduli spaces, exemplifying with the $Y^{p,q}$ spaces.

%
%
%
\section{Explicit Expressions for Plethystics}\label{s:explicit}\setequation
With the plethystic programme thus outlined above, it is expedient to
present some useful results concerning the generating
functions $f$ and $g$.
First, let us take a closer look at the fundamental
relation of the plethystic inversion formula:
\bea \nn
g(t) &=&
PE[f(t)] :=
PE[\sum_{k=0}^\infty a_k t^k] =
\exp\left[ \sum_{p=1}^\infty \frac{1}{p} (f(t^p) - f(0))
  \right] =
\prod_{m=1}^{\infty} \frac{1}{(1 - t^m)^{a_m}}
\Leftrightarrow \\
\label{PE-invert}
f(t)-f(0) &=& PE^{-1}[g(t)] = \sum_{l=1}^\infty \frac{\mu(l)}{l} \log(
g(t^l)) \ .
\eea
The above expression is a central motif for the plethystic programme
and the proof of which was not presented in \cite{Benvenuti:2006qr},
nor, for that matter, could one find it, within a body of literature
often obscured by mathematical sophistry, in an explicit
fashion. The proof is, in fact, rather straight-forward, which we shall
presently see.

Taking the logarithm of the product form of PE in \eref{PE-invert}
and series-expanding, we have
\beq\label{log-g}
\log(g(t)) = \sum_{k=1}^\infty  (-a_k) \sum_{m=1}^\infty
-\frac{1}{m}(t^{k})^m \ .
\eeq
Whence,
\bea\nn
PE^{-1}[g(t)] =
\sum_{l=1}^\infty \frac{\mu(l)}{l} \log( g(t^l)) &=&
\sum_{l=1}^\infty \frac{\mu(l)}{l}
\left( \sum_{k=1}^\infty  a_k \sum_{m=1}^\infty \frac{1}{m}(t^{lk})^m
\right) \\ \label{fg-invert}
&=& \sum_{k=1}^\infty a_k \sum_{n=1}^\infty \sum_{l|n} \mu(l)
\frac{1}{n} (t^k)^n \ ,
\eea
where we have re-written the double sum on $m$ and $l$ as the
alternative sum on $n = m\ l$ and its divisors $l$.
Using a fundamental theorem of analytic number theory, viz., the
M\"obius inversion formula \cite{serre}
\[
\sum_{d|n} \mu(d) = \delta_{n,1} \ ,
\]
the double sum $\sum\limits_{n=1}^\infty \left(
\sum\limits_{l|n} \mu(l) \right)
\frac{1}{n} (t^k)^n $ simply reduces to $t^k$, whereby making the RHS
of \eref{fg-invert} equal to $\sum\limits_{k=1}^\infty a_k t^k = f(t) -
f(0)$, as is required.

Next, the $\nu$-inserted version of PE is of vital importance:
\beq\label{g-nu}
g(\nu, t) = \prod_{m=0}^{\infty} \frac{1}{(1 - \nu  \, t^m)^{a_m}}
= \sum_{N=0}^\infty g_N(t) \nu^N \ .
\eeq
This simple insertion gives us, almost miraculously,
the powerful generating functions $g_N$ which capture the
multi-trace GIO's for any finite $N$ and from which the counting $f_N$
for the single-trace GIO's can be extracted by the plethystic
logarithm, i.e., $f_N = PE^{-1}[g_N(t)]$. The
remarkable fact is that $g_N(t)$ requires only the knowledge of the
Hilbert series $f(t) := f_\infty(t) = \sum\limits_{m=0}^\infty a_m
t^m$, which we recall from our outline above, is the fundamental
object obtained purely from the geometry of the Calabi-Yau singularity
$\cM$. Explicit expressions fot $g_N$, especially its large-$N$
behaviour, are certainly important in, for example, entropy-counting
of bulk black-hole states.

\subsection{All $g_N$ as Functions of $g_1$}
Now, from the series expansion \eref{g-nu},
we can find recursion relations among the coefficients of expansion,
whereby expressing our desired $g_N$ in terms of the basic Hilbert
series $g_1=f_\infty$.
As an enticement, for example, we notice that:
\bean
{\partial^2 g(\nu, t) \over \partial \nu^2} & = &
\left(\sum_{k=0}^\infty \frac{a_k t^k}{(1 - \nu \, t^k)}\right)^2
g(\nu, t) +
g(\nu, t) \sum_{k=0}^\infty \frac{a_k t^{2k}}{(1 - \nu \, t^k)^2} \ , \\
{\partial^3 g(\nu, t) \over \partial \nu^3} & = &
\left(\sum_{k=0}^\infty
\frac{a_k t^k}{(1 - \nu \, t^k)}\right)^3 g(\nu, t) + 3
g(\nu, t)\left(\sum_{k=0}^\infty \frac{a_k t^k}{(1 - \nu \,
t^k)}\right)\left(\sum_{k=0}^\infty \frac{a_k t^{2k}}{(1 - \nu \,
t^k)^2}\right)+ \\
&& \qquad  + g(\nu, t)\left(\sum_{k=0}^\infty \frac{2 a_k t^{3k}}{(1 -
\nu \, t^k)^3}\right) \ .
\eean
From this we have
\bea
\nn 
g_{2}(t) & = &
\frac{1}{2!} {\partial^2 g\over \partial \nu^2}|_{\nu=0} =
{1\over 2} [ g_{1}^2(t)+ g_{1}(t^2)] \ ,
\\
g_{3}(t) & = &
\frac{1}{3!} {\partial^3 g\over \partial \nu^3}|_{\nu=0} =
{1\over 6} [g_{1}^3(t)+ 3 g_{1}(t)g_{1}(t^2)+ 2 g_{1}(t^3)] \ .
\label{gNg1}
\eea

\paragraph{A Systematic Approach: }
We can obtain the above results more systematically. Recalling that
the fundamental definition of PE has two equivalent expressions, as
a sum or as a product (q.v.~\eref{PE-invert}), we have that
\beq\label{g-nu-PE} 
g(\nu,t) = \prod_{m=0}^{\infty} \frac{1}{(1 -
\nu  \, t^m)^{a_m}} = \exp\left( \sum_{k=1}^\infty \frac{1}{k}
g_1(t^k) \nu^k  \right) \ , \eeq where $g_1(t) = f_\infty(t) =
\sum\limits_{m=0}^{\infty} a_m t^m$.
Hence,
\[
\sum_{N=0}^\infty g_N(t) \nu^N = \exp\left(
\sum_{k=1}^\infty \frac{1}{k} g_1(t^k) \nu^k
\right) \ .
\]
Expanding the exponential in the RHS gives a series in powers of
$\nu$:
\bean
g(\nu,t) &=&
1 + g_1(t)\,\nu + \frac{\left( {g_1(t)}^2 + g_1(t^2) \right) \,{\nu}^2}{2} + 
  \frac{\left( {g_1(t)}^3 + 3\,g_1(t)\,g_1(t^2) + 2\,g_1(t^3) \right)
       \,{\nu}^3}{6} + \\
&& +
\frac{\left( {g_1(t)}^4 + 6\,{g_1(t)}^2\,g_1(t^2) + 
       3\,{g_1(t^2)}^2 + 8\,g_1(t)\,g_1(t^3) + 6\,g_1(t^4) \right) \,
     {\nu}^4}{24} + \cO(\nu^5)
\eean
Thus, very straight-forwardly, we obtain the expressions for $g_N(t)$
by simply reading off the coeffcients of $\nu^N$; giving us the
desired generating function $g_N(t)$ in terms of the Poincar\'e series
$g_1$ with powers of its argument $t$. The results for $N=2,3$ are
seen to agree with those in \eref{gNg1}.

\subsection{Relation to Young Tableaux}
\begin{figure}
\centerline{\epsfxsize=4in\epsfbox{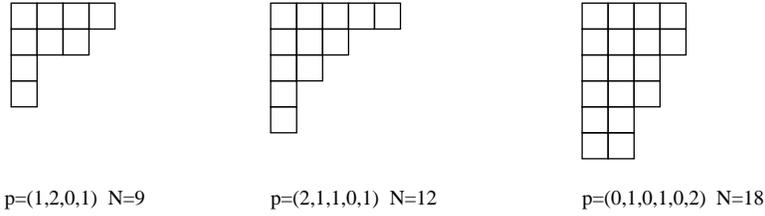}} 
\caption{{\sf Examples of Young
Tableaux with partition $p=\{p_1,p_2,p_3,...,p_k,...\}$ and $N$.
The constraint is that $N = \sum\limits_{k=1} p_k k$.
}}
\label{f:Young}
\end{figure}

One can proceed further with the above expansion for $g_N$, 
and obtain interesting connections to Young
tableaux. From \eref{g-nu-PE}, one can series-expand the exponential as:
\beq\label{gN}
\sum_{N=0}^\infty g_N(t) \nu^N = \exp\left(
\sum_{k=1}^\infty \frac{1}{k} g_1(t^k) \nu^k
\right)=\prod_{k=1}^\infty e^{\left( \nu^k {g_1(t^k)\over k}\right)}
= \prod_{k=1}^\infty  \left( \sum_{p_k=0}^\infty  \nu^{k p_k}
\frac{g_1(t^k)^{p_k}}{p_k! \ k^{p_k}} \right) \ .  
\eeq
Now, which terms contribute to $\nu^N$? We see that this is whenever
\beq\label{Division} 
\sum_{k=1}^\infty p_k k = N \ . 
\eeq
Under this constraint we have the explicit expression for $g_N(t)$ as
\beq\label{coeff-Division} 
g_N(t) =
\sum\limits_{{\tiny \ba{l} p_1,p_2,..\\
  \sum\limits_{k=1}^\infty p_k k= N \ea}}
\prod_{k=1}^\infty
\frac{(g_1(t^k))^{p_k}}{p_k! \ k^{p_k}} \ . 
\eeq
The relation \eref{Division} is a familiar combinatorial problem: 
the partition of $N$ into increasing components $k=1,2,3\ldots$ 
of respective multiplicity
$p_k$. This is, of course, just the Young Tableau; to see it we
just draw $p_k$ columns of length $k$ from right to left with $k$
increasing. For clarity, we have drawn a few illustrations in \fref{f:Young}
with given $p=\{p_1,p_2,p_3,...,p_k,...\}$. For example, for the first
tableau, there is a total of 9 boxes. The vector $(1,2,0,1)$ means that
$p_1=1, p_2=2, p_3=0$ and $p_4=1$.
Now, $p_1=1$ means that there is 1 column with only one box; 
this is the first column
from the right. Similarly, there are $p_2 = 2$ columns with 2 boxes
and $p_3=0$ means there are no columns with 3 boxes. Finally, $p_4=1$
means there is one column with 4 boxes, the one to the far left. 

We wish to emphasize that the natural emergence of Young Tableaux is not
an accident and has deep connections to Hilbert scheme which we will
touch upon later. The reader is referred to the recent works of \cite{japan}.
At a superficial level, we have related each term in the sum 
\eref{coeff-Division}
to a given Young Tableau. In other words, given a Young Tableau we can
count the number of columns with length $k$, say it is $p_k$; then
we can assign one factor $\frac{(g_1(t^k))^{p_k}}{p_k! \ k^{p_k}}$.
Multiplying all factors together we get contribution for the
particular Young Tableau. Finally we sum up all Young Tableaux with
box number $N$, giving us the $g_N$ we need.

\paragraph{A Fermionic Version?}
As a brief digression,
one notices that the expression for the plethstic exponential, in its
product form, is a generating function for a bosonic oscillator.
One might wonder what the fermionic counter-part
signifies. In other words, we could define, for $f(t) =
\sum\limits_{n=0}^\infty a_n t^n$,
\[
\W{PE}[f(t)] := \prod\limits_{k=1}^\infty (1+t^k)^{a_k}, \qquad
\W{PE}_\nu[f(t)] := \prod\limits_{k=0}^\infty (1+ \nu t^k)^{a_k} \ .
\]
It would be interesting to find what these may count in the D-brane
gauge theory and what nice inverse functions they possess.

\subsection{Generalising Meinardus}\label{s:asy}
The asymptotic expressions for the generating functions are clearly of
importance. In \cite{Benvenuti:2006qr}, we discussed at length the
so-called Meinardus theorem \cite{meinardus} 
which generalises the Hardy-Ramanujan
formula for the partition of integers and gives the asymptotics of the
function $g_\infty(t)$. 
Now, what about the aymptotic expressions of $g_N(t)$ where we have a
finite number $N$ of D3-branes? In other words, we wish to know,
as $n \to \infty$ in the expansion 
\beq\label{gN-expand}
g_N(t) = \sum_{n=0}^\infty g_N(n) t^n \ ,
\eeq
the behaviour of $g_N(n)$ for a given $N$.

Thus, we need a generalisation of Meinardus to include
$\nu$-insertions. Luckily, there is a result due to
Haselgrove-Temperley \cite{HT} with certain relaxation of conditions
in \cite{rich}. The fermionic version mentioned above has its asymptotics
studied in detail by \cite{hwang}. The key result of \cite{HT,rich}
is, under certain convergence conditions into which we shall not
delve, that
\begin{theorem}\label{asymp}
For $G(\nu, t) = \prod\limits_{r=1}^\infty (1 - \nu
t^{\lambda_r})^{-1} = 
\sum\limits_{n,N=0}^\infty g_N(n)t^n \nu^N$, define 
\[\ba{rcl}
\Psi(x) &:=& \log G(x) = - \sum\limits_{r=1}^\infty \log(1 - e^{x
  \lambda_r}),\\
K(x) &:=& \prod\limits_{r=1}^\infty \left(1 + \frac{x}{\lambda_r}\right)^{-1}
  e^{x/\lambda_r}, \quad
F(y) := \frac{1}{2\pi i} \int\limits_{-i \infty}^{i \infty}
  K(x) e^{xy} dx, \\
\xi & := & \mbox{ a root of } \Psi'(\xi) + n = 0 \ , \quad
N_0 := \sum\limits_{r=1}^\infty (e^{\xi \lambda_r}-1)^{-1}, \\
\ea\]
then, the asymptotics (for $n$ large and $N$ fixed) are:
\[
g_N(n) \sim \xi F((N-N_0)\xi) \ g(n), \qquad
g(n) \sim \left( 2\pi \Psi''(\xi) \right)^{-\frac12}
e^{\Psi(\xi) + n \xi} \ .
\]
\end{theorem}
Of course, we need to recast our $g(\nu, t)$ in \eref{g-nu} into the
form which the theorem addresses; this is a redefinition of the
$\lambda_r$ in terms of the $a_m$ to eliminate repetitions:
\beq\label{a2lam}
\lambda_r = \left\{\ba{ccl}
1 && r = a_0, \ \ldots, \ a_1;\\
2 && r = a_1 + 1, \ \ldots, \ a_1 + a_2;\\
3 && r = a_1 + a_2 + 1, \ \ldots, \ a_1 + a_2 + a_3;\\
\ldots
\ea\right.
\eeq

We see that the function $G(x) = \exp(\Psi(x))$ is when the
$\nu$-insertion is absent (note that here counting does start from
$r=1$) and should capture the original Meinardus
result for the plethystic exponential. 
Importantly, a key property of $\Psi(x)$, in terms of the
$a_m$ coefficients (cf.~\cite{actor}), is that its asymptotic
behaviour is
\beq\label{psi-x}
G(x) = e^{\Psi(x)} = \prod\limits_{r=1}^\infty (1 - e^{x r})^{-a_r}
\sim
\exp\left[
A \Gamma(\alpha) \zeta(\alpha+1) x^{-\alpha} - D(0) \log x + D'(0)
\right] \ ,
\eeq
where $D(s) := \sum\limits_{m=1}^\infty \frac{a_m}{m^s}$ is the
Dirichlet series which has only 1 simple pole at $s = \alpha \in
\IR_+$ with residue $A$.

Using \eref{psi-x} and its derivative,
we see that the quantities $\xi$ and $g(n)$ in
Theorem \ref{asymp} explicitly evaluate to, for large $n$,
\bea
\nn \xi & \sim & \mbox{Root}\left[
-A\Gamma(\alpha+1)\zeta(\alpha+1)x^{-\alpha-1} - D(0)/x + n = 0
\right] \sim \left(\frac{1}{n} A \Gamma(\alpha+1) \zeta(\alpha+1)
\right)^{\frac{1}{\alpha+1}} \\
\nn g(n) & \sim &
C_1 n^{C_2} \exp
\left[ n^{\frac{\alpha}{\alpha+1}}(1+\frac{1}{\alpha})\left(A
\Gamma(\alpha+1) \zeta(\alpha+1)\right)^{\frac{1}{\alpha+1}}
\right] \\
\label{xi-gn}
C_1 &:=& e^{D'(0)} \frac{1}{\sqrt{2\pi(\alpha+1)}} \left(
    A \Gamma(\alpha+1)
    \zeta(\alpha+1)\right)^{\frac{1-2D(0)}{2(\alpha+1)}}, \quad
C_2 := \frac{D(0)-1-\frac{\alpha}{2}}{\alpha+1} \ .
\eea
We see that $g(n)$ above is exactly the Meinardus result
\cite{meinardus,actor}
for the asymptotics of the plethystic exponential without
$\nu$-insertion (cf.~also, Section 6 of \cite{Benvenuti:2006qr}). 
In other words, the content of Theorem \ref{asymp} is
that the pre-factor
\beq\label{prefactor}
\xi F\left( (N-N_0) \xi \right)
\eeq
encodes the effects of $\nu$-insertion, i.e., the $N$-dependence,
to the classical Meinardus asymptotic formula for $g(n)$ in
\eref{xi-gn}.
For values of $n<N$ the expression for $g_N(n)$ should coincide
precisely with that of $g_\infty(n)$ as the pre-factor tends to 1.
On the other hand, for $n>N$ there will be corrections and the $g_N(n)$
is expected to be smaller than $g_{\infty}(n)$; this is because the
counting should be less at finite $N$ since there are constraints
which vanish at infinite $N$.

\paragraph{Example: $\IC$  }
Let us first check a simple case. Let $a_m = 1$ for all $m \in \IZ_{
\ge 0}$. This is where the Hilbert series is equal to $f_\infty(t) =
(1-t)^{-1}$ and we recall \cite{Benvenuti:2006qr} that  the geometry
is just $\IC$. The conversion
\eref{a2lam} makes $\lambda_r = r$, which is a specific example
considered on p238 of \cite{HT}, giving us
\[
K(x) =  \prod\limits_{r=1}^\infty \left(1 + \frac{x}{r}\right)^{-1}
  e^{x/r} = e^{\gamma x} \Gamma(x+1), \qquad
F(y) = \exp\left( - (\gamma + y) - e^{-(\gamma + y)}\right) \ , 
\]
where $\gamma := \lim\limits_{n \to \infty} \left(
\sum\limits_{j=1}^n j^{-1} - \log(n) \right)$ is the Euler
constant\footnote{
  Indeed, we can see this since $F(y) =
  \sum\limits_{n=-1,-2,-3,\ldots} \res\limits_{z \to n}
  \Gamma(z+1)e^{\gamma z + z y} =  \sum\limits_{n=0}^\infty
  \frac{(-1)^{n}}{n!} e^{-(n-1)(\gamma + y)} = e^{-1/a}/a$,
  for $a = \exp(\gamma+y)$.
  }.
The Dirichlet series is here $D(s) = \sum\limits_{m=1}^\infty m^{-s} =
\zeta(s)$; whence $\alpha = A = 1$ with $D(0) = -\frac12$ and $D'(0) =
-\frac12 \log(2\pi)$. Therefore, \eref{psi-x} dictates that
\[
\Psi(x) \sim \frac{\pi^2}{6x} + \frac12 \log x - \frac12\log{2\pi} 
\Rightarrow
\Psi'(x) \sim -\frac{\pi^2}{6x^2} + \frac{1}{2x}, \
\Psi''(x) \sim \frac{\pi^2}{3x^3} - \frac{1}{2x^2}
\ .
\]
By \eref{xi-gn} we thus have
\beq\label{gn-C}
\xi \sim \frac{-3+\sqrt{24n \pi^2 + 9}}{12n} \sim \frac{\pi}{\sqrt{6n}}
\ , \quad
g(n) \sim \left( 2\pi \Psi''(\xi) \right)^{-\frac12}
e^{\Psi(\xi) + n \xi} \sim \frac{1}{4\sqrt{3} n} e^{\pi \sqrt{2n/3}} \ .
\eeq
Indeed, $g(n)$ is exactly the famous Hardy-Ramanujan asymptotic behaviour
for the $\eta$-function. The effect of the $\nu$-insertion is then
apparent in the pre-factor governed by the function $F$. Now, we see
that, for small $x$,
\beq\label{n0-lam=r}
N_0(x) = \sum\limits_{r=1}^\infty (\exp(r x)-1)^{-1}
\sim \sum\limits_{r=1}^{1/x} \frac{1}{r x} +
\sum\limits_{r=1/x}^{\infty} \exp(-r x) 
=
-\frac{H(1/x)}{x} + \frac{e^{x-1}}{e^x-1}
\sim -\frac{\log x}{x}
\ ,
\eeq
where we have series-expanded for the first part of the sum ($H(x)$ is
the Harmonic number) and
neglected the small contribution of the $-1$ in the denominator for
the second sum. Therefore, since $n$ is large, we can apply
\eref{n0-lam=r} to give us
\[
N_0 = N_0(\xi) \sim 
\frac{\sqrt{6n}}{\pi} \log \frac{\sqrt{6n}}{\pi}  \ .
\]
Thus, we can write the pre-factor in \eref{prefactor}
($n$ is large and $N$ is fixed) as
\[\ba{rcl}
\xi F \left( (N-N_0) \xi \right) 
& \sim & \frac{\pi}{\sqrt{6n}}
\exp\left( - (\gamma + (N-N_0)\frac{\pi}{\sqrt{6n}}) - 
e^{-(\gamma + (N-N_0)\frac{\pi}{\sqrt{6n}})}\right) \\
& \sim & 
\frac{\pi}{\sqrt{6n}}\exp\left( \log \frac{\sqrt{6n}}{\pi}  - \frac{N
  \pi}{\sqrt{6n}} - 
e^{\log \frac{\sqrt{6n}}{\pi}  - \frac{N \pi}{\sqrt{6n}} }
\right) \sim
\exp(- \frac{N \pi}{\sqrt{6n}} -
   \frac{\sqrt{6n}}{\pi} e^{- \frac{N \pi}{\sqrt{6n}} }) 
\ .
\ea\]
\comment{
It is re-assuring that for $N \to \infty$, this pre-factor goes to 1,
as it should, since $g_\infty(n)$ is the classical Meinardus 
result without the $\nu$-insertion, as shown in \cite{Benvenuti:2006qr}.
}
In summary, we have the asymptotic expansion of $g_N(n)$ as
\beq
g_N(n; \IC)
\sim 
\frac{1}{4\sqrt{3}n}
\exp\left( \pi \sqrt{\frac{2n}{3}} \right)
\exp \left[- \frac{N \pi}{\sqrt{6n}} -
   \frac{\sqrt{6n}}{\pi} e^{- \frac{N \pi}{\sqrt{6n}}}
\right]
\ .
\eeq
We have actually reproduced a classical result of \cite{ACG}, which is
also studied recently in Bose-Einstein condensates in
\cite{GH}. Specifically, the above result agrees completely with Eq.(13) of
\cite{GH}, wherein they have simplified the expression to
$\frac{g(n)}{\sqrt{n}} \exp(-\frac{2}{c} \exp(x_N(n)) - x_N(n))$ with
$c = \sqrt{\frac23}\pi$, $g(n)$ given in \eref{gn-C} and $x_N(n) :=
\frac{cN}{2\sqrt{n}} - \log(\sqrt{n})$.

\subsection{A Large Class of Examples}
Thus emboldened, we may proceed to more examples.
Since the plethystic exponential has a singularity at $t = 1$ at $\nu
= 1$, it is expedient to study contributions of the form
\beq\label{f-1-t}
f(t) = f_\infty(t;\cM) = g_1(t;\cM) =
\frac{V_3}{(1-t)^{3}} + \frac{V_2}{(1-t)^{2}} + \frac{V_1}{1-t}
+ V_0 + \cO(1-t)
\ ;
\eeq
we go up to poles of order
3 because $\cM$ is at most 3-dimensional in the cases
of concern.
Physically, $V_3$ can be thought of as the volume of the
dual AdS horizon, i.e., the normalised volume of the Sasaki-Einstein
manifold
(cf.~\cite{Martelli:2006yb,Martelli:2006vh,Butti:2006nk,Butti:2006au}),
and $V_i$ are related to the components of the Reeb vectors.

It turns out, for what we shall shortly describe in the next
section, that we do not need as refined an attack as
Haselgrove-Temperley, but, rather, a leading order analysis. Indeed,
the results of \cite{HT} for $d > 1$ require a regularisation into whose
subtleties we presently do not wish to venture. We shall, instead,
follow the saddle-point method in the physics literature, such as 
\cite{BEC}.  Indeed we are essentially studying a contour
integral
\[
g_N(n) = \frac{1}{(2\pi i)^2} \oint_{\Gamma_{\nu=0}} d\nu
\oint_{\Gamma_{t=0}} dt \frac{g(\nu,t)}{\nu^{N+1} t^{n+1}} \ ,
\]
which picks up the
residues at the poles and the form in \eref{f-1-t} will be dominant. 
The statement, with the
same notations as above, is as follows. For both $N$ and $n$ large
(note that Haselgrove-Temperley only requires that $n$ be large),
\bea\label{saddle}
\nn 
&&g_N(n) \sim g(\nu_0, t_0) \nu_0^{-N-1} t_0^{-n-1}, \quad \text{where}
\\ 
&&\left[ N+1 = \nu \frac{\partial}{\partial \nu} \log g(\nu,t)
  \right]_{\nu_0, t_0} \ ,
\left[ n+1 = t \frac{\partial}{\partial t} \log g(\nu,t)
  \right]_{\nu_0, t_0} \ .
\eea

We can first directly evaluate $\log g(\nu,t)$. From \eref{f-1-t}, we
have that
\bean
{\hspace{-1in}
a_n }&=& 
V_3 \frac{(n+1)(n+2)}{2} + V_2(n+1) + V_1 + V_0 \delta_{n,0}
\qquad \Rightarrow\\
{\hspace{-1in}
\log g(\nu,t)} &=&  -\sum_{n=0}^\infty a_n \log (1-\nu t^n) \\
&=&  -V_0 \log(1-\nu) +
 \sum_{k=1}^\infty{\nu^k\over k} \left[
\frac12 V_3 Li_{-2}(t^k) +
(V_2 + \frac32 V_3) Li_{-1}(t^k) + 
(V_1+V_2+V_3)(1+Li_{0}(t^k))
\right]
\ ,
\eean
where we have used the definition of the Polylogarithmic function
$Li_d(x) = \sum\limits_{n=1}^\infty {x^n\over n^d}$. In fact, for
$d \in \IZ_{\le 0}$, these are simply rational functions.

Recall now that we wish to study the behaviour of $g(\nu,t)$ near
$t=1$ and $\nu =1$. 
Hence, we can define $t:=e^{-q}$ and $\nu = e^{-w}$ and will
study the behaviour near $q,w\to 0$.
Series expanding $\log g(\nu,t)$ and keeping dominant contributions in
the inverses of $q$ and $w$, we find that
\bea 
\log g(w,q) 
\nn& \sim & \sum_{k=1}^\infty {\nu^k\over k}
\left[ V_{0} + \frac{V_{1}}{2} + \frac{5\,V_{2}}{12} + 
  \frac{3\,V_{3}}{8} + 
  \frac{V_{1} + V_{2} + V_{3}}{k\,q} + 
  \frac{2\,V_{2} + 3\,V_{3}}{2\,k^2\,q^2} + \frac{V_{3}}{k^3\,q^3} 
\right]\\
&\sim& \frac{V_3}{q^3}(\zeta(4) - \zeta(3) w) - 
(V_{0} + \frac{V_{1}}{2} + \frac{5\,V_{2}}{12} + \frac{3\,V_{3}}{8}) \log(w)
\ . \label{logg-nu-t}
\eea

We are now ready to solve for the saddle points given in
\eref{saddle}. Since $t=e^{-q}, \nu = e^{-w}$, we have
$t{\partial \over \partial t}=-{\partial \over \partial q}$ and
$\nu{\partial \over \partial \nu}=-{\partial \over \partial w}$ and
the saddle equations read:
\bean
n+1 &=& -{\partial \log g(w,q)\over \partial q} \sim 
3\zeta(4)V_3 q^{-4} \ ,
\\
N+1 &=&  -{\partial \log g(w,q)\over \partial w} \sim
\zeta(3) V_3 q^{-3} + (V_{0} + \frac{V_{1}}{2} + \frac{5\,V_{2}}{12} +
\frac{3\,V_{3}}{8}) w^{-1} \ .
\eean
Therefore, the saddle points are
\beq\label{saddle-sol}
q_0 \sim \left( \frac{3V_3\zeta(4)}{n}\right)^{\frac14} \ , \qquad
w_0 \sim (V_{0} + \frac{V_{1}}{2} + \frac{5\,V_{2}}{12} +
\frac{3\,V_{3}}{8}) \left(
   N - \zeta(3) V_3 q_0^{-3}
\right)^{-1}
\eeq
These results are encouraging. For $V_{0,1,2}=0$ and $V_3=1$, the case
was studied in nice detail in \cite{BEC}. The expressions in
\eref{saddle-sol}, to leading order, agree exactly with their
Eq. (17-19), in {\it cit. ibid.}
Substituting back into \eref{saddle}, we conclude that, to leading
order,
\bea
\nn \log g_N(n) &\sim& \log g(\nu_0, t_0) + N w + n q \\
\ &\sim& C_0 n^{\frac34} +
C_1 \left[
\frac{N}{N -  C_2 n^{\frac34}}
+ \log\left(N - C_2 n^{\frac34} \right)
\right] \ ; \\
\nn 
C_0 & := & 3^{-\frac34} 4 (V_3\zeta(4))^{\frac14}, \quad
C_1 := V_{0} + \frac{V_{1}}{2} + \frac{5\,V_{2}}{12}
+\frac{3\,V_{3}}{8}, \quad 
C_2 := \zeta(3)V_3^{\frac14} (3\zeta(4))^{-\frac34} \ .  
\eea
Once more, we are re-assured. The first term, which only depends on
$n$, should be the classical Meinardus result while the second is the
pre-factor \eref{prefactor} discussed above. We have done the
Meinardus analysis for $\IC^3$ in \cite{Benvenuti:2006qr};
substituting $\zeta(4) = \frac{\pi^4}{90}$ gives us the first term as
$\frac{2\cdot 2^{\frac34} \pi}{3 \cdot 15^{\frac14}} n^{\frac34}$,
precisely the exponent of $p_1$ in Eq (6.7) of \cite{Benvenuti:2006qr}.

Another interesting limit to consider is $t \sim 1$ and $\nu \sim 0$. 
Here, we expand about $q = -\log t$ and $\nu$ directly and
\eref{logg-nu-t} becomes
\beq
\log g(\nu,t) \sim \nu V_3 q^{-3} \ ,
\eeq
giving us the saddle points $q_0 \sim 3N/n$ and $\nu_0 \sim
(3N/n)^3N/V_3$. Thus,
\beq\label{small-nu}
\log g_N(n) \sim 4N\left(\log(N^{-1} n^{\frac34}) + 1 -\frac14 \log
\frac{27}{V_3}
\right) \ .
\eeq
Note that in order for $\nu \sim 0$, we need $N \ll n^{\frac34}$.
%
%
\subsection{The Entropy of Quiver Theories}\label{s:entropy}
Having expounded upon a collection of examples and demonstrated the
explicit power of the Halselgrove-Temperley result as well as
saddle-point evaluations in generalising
Meinardus, let us now address a problem of great physical interest. 
A chief motivation for finding explicit expressions, in paricular the
asymptotic behaviour, of our generating functions is to determine the
number of degrees of freedom, i.e., the entropy of the gauge theory. 
Indeed, as the Hardy-Ramanujan formula
is central to the determining the entropy of the bosonic critical
string, the results presented in the previous section will be
essential to that of D-brane probe theories.

The growth of the number of our mesonic BPS operators in the gauge
theory can be a good estimate for the entropy of the system. More
generally, it serves as a lower bound for the total number of
operators in the gauge theory, regardless of whether they are BPS or
not. Thus if we are looking for an underlying black hole entropy,
the discussions above will be greatly pertinent.
Specifically, in our context of the gauge theory of $N$ D-branes
probing a geometry $\cM$, we can define the entropy 
$S_N(n)$ as
\beq
S_N(n) = \log g_N(n) 
\quad \mbox{where} \quad
g(\nu, t; \cM) := \sum_{N,n=0}^\infty g_N(n) t^n \nu^N \ ,
\eeq
and we recall that $g(\nu, t; \cM)$ is the $\nu$-inserted plethystic
exponential of the Hilbert series (the fundamental generating function
$f$) of the geometry of $\cM$.

Now, we would like to compute critical exponents depending on
dimensionality. Therefore, we need to consider the generalisation of
\eref{f-1-t} to
\beq
f(t) = \frac{V_d}{(1-t)^{d}} + \ldots +
\frac{V_2}{(1-t)^{2}} + \frac{V_1}{1-t} + V_0 + \cO(1-t) \ .
\eeq
Following the computation performed above, we easily see that the
saddle points are (in the $t, \nu \sim 1$ limit) now
\beq
q_0 \sim \left( \frac{dV_d\zeta(d+1)}{n}\right)^{\frac{1}{d+1}} 
\ , \qquad
w_0 \sim \left[V_0 + \sum_{j=1}^d V_j (1+\sum_{i=0}^{j-1} \beta_i
  \zeta(-i))\right]
\left(
   N - \zeta(d) V_d q_0^{-d}
\right)^{-1} \ , 
\eeq
where $\beta_i$ are coefficients such that 
\beq
{n+d-1 \choose {d-1}} := \sum_{i=0}^{d-1} \beta_i n^i \ .
\eeq
Substituting into the saddle point equation, we find the entropy to be 
\beq
S_N(n) \sim C_0 n^\alpha + C_1 \left[
\frac{N}{N -  C_2 n^{\alpha}} + \log\left(N - C_2 n^{\alpha}\right)
\right] \ ,
\eeq
where the critical exponent is $\alpha = \frac{d}{d+1}$ 
and the constants are
\bean
C_0 &=& d^{-\frac{d}{d+1}}(d+1)(V_d\zeta(d+1))^{\frac{1}{d+1}} \ , \\
C_1 &=& V_0 + \sum_{j=1}^d V_j (1+\sum_{i=0}^{j-1} \beta_i \zeta(-i)) \ , \\
C_2 &=& \zeta(d)V_d^{\frac{1}{d+1}} (d\zeta(d+1))^{-\frac{d}{d+1}} \ .
\eean
We remark, upon obtaining a similar expression as \eref{small-nu}
for $\nu \sim 0$, that
our treatment gives rise to a critical regime in which there is a
cross over between $\nu \sim 0$ and $\nu \sim 1$. 
This critical regime is given
by the order parameter $N \sim n^{d/d+1}$ or, alternatively,
$n \sim  N^{1+1/d}$. 
When the two sides are of the same order we are in the $\nu \sim 1$
regime and the number of operators is controlled by $n$ essentially.
When the order parameter is small the number of operators depends on
$N$. 
%
\section{$SU(2)$ Subgroups: ADE Revisited}\label{s:su2}\setequation
We have, in the above, discussed extensively the various general
properties of the generating functions, the recursions, 
relations to Young tableaux,
and especially the asymptotics. Now, let us move on to some specific
examples.
An extensively studied class of CY singularities are orbifold
theories. Of particular mathematical interest has been the local-K3
singularities, viz., $\IC^2 / G$ where $G$ is a discrete, finite
subgroup of $SU(2)$. Such groups fall under an ADE-pattern and the
quivers are central to the McKay Correspondence.

In \cite{Benvenuti:2006qr}, we computed the fundamental generating
functions, i.e., the Hilbert series $g_1 = f_\infty$. We recall that
for orbifolds of finite group $G$, the Hilbert series is computed by
the so-called Molien series
\beq\label{molien}
f_\infty(t; G) = M(t;G) = \frac{1}{|G|}\sum_{g \in G}
\frac{1}{\det(\II - t g)} \ .
\eeq
A natural
question to ask is what explicit expressions can be derived for
$g_N$ at finite $N$. Using the prescription in the previous section,
we can readily expand a few terms of \eref{gN} to see what we obtain.
Take the example of $G = \hat{D}_4$, the binary dihedral group of
order 8, which was investigated in detail in \cite{Benvenuti:2006qr},
the Hilbert series is the Molien series
\beq\label{g1-D4}
g_1(t) = M(t;~\hat{D}_4) = \frac{1 + t^6}
{{\left( 1 - t^4 \right) }^2} \ .
\eeq
Substituting \eref{g1-D4} into \eref{gN}, we obtain
($g_0(t) = 1$ automatically):
\[
g_2(t) =
\frac{1 - t^2 + t^6 + t^8 - t^{12} + t^{14}}
  {\left( 1 - t^2 \right) \,{\left( 1 - t^4 \right) }^2\,\left( 1 -
    t^8 \right) } \ , \qquad
g_3(t) = 
\frac{1 - t^2 + 2\,t^8 + t^{12} + t^{18} + 2\,t^{22} - t^{28} + t^{30}}
  {\left( 1 - t^2 \right) \,{\left( 1 - t^4 \right) }^2\,
    \left( 1 - t^6 \right) \, \left( 1 - t^8 \right)\, \left( 1 -
    t^{12} \right) } \ , \quad
\ldots
\]
We see that these coefficients quickly become
complicated. Nevertheless, the algorithm is clear and one may extract
$g_N$ {\it ad libertum}.

\subsection{Recursion Relations and Difference Equations}
Let us entice the reader with some immediately noticeable curiosities
for the series-coefficients for the Hilbert (Molien) series for the ADE
orbifolds. Take 
the A-family (where $\hat{A}_{n-1} := \IZ_n$). We recall from
\cite{Benvenuti:2006qr} that
\[
f_\infty(t;~\hat{A}_{n-1}) = \frac{(1+t^n)}{(1-t^2)(1-t^n)} \ .
\]
We see that
\beq\label{someAexp}
\ba{rcl}
f_\infty(t;~\hat{A}_1) 
&=& 1 + 3\,t^2 + 5\,t^4 + 7\,t^6 + 9\,t^8 + 11\,t^{10} + 
  13\,t^{12} + 15\,t^{14} + 17\,t^{16} + 19\,t^{18} + 
  21\,t^{20} + \cO(t^{21})\\
f_\infty(t;~\hat{A}_3) &=& 1 + t^2 + 3\,t^4 + 3\,t^6 + 5\,t^8 + 5\,t^{10} + 
  7\,t^{12} + 7\,t^{14} + 9\,t^{16} + 9\,t^{18} + 
  11\,t^{20} + \cO(t^{21})\\
f_\infty(t;~\hat{A}_5) 
&=& 1 + t^2 + t^4 + 3\,t^6 + 3\,t^8 + 3\,t^{10} + 
  5\,t^{12} + 5\,t^{14} + 5\,t^{16} + 7\,t^{18} + 
  7\,t^{20} + \cO(t^{21})
\ea\eeq
Thus, for $n=2k$ even, the pattern of the coefficients is
$\{1,\ldots,1;~3,\ldots,3;~5,\ldots,5;\ldots \}$.

In fact, we will now proceed to find analytic expressions for the
series-coefficients, i.e., the number of single-trace GIO's, of
$f_\infty$ for all the discrete, finite subgroups of $SU(2)$. This
indeed places our generating function in full power and provide us
with invariants of arbitrary degree immediately. The reason we can do
so is because the Molien series is a rational function in $t$ and
indeed, for any rational function, one could systematically obtain
recursion relations, which can then be solved.
It is easiest to start with the exceptionals, i.e., the E-family, with
which we shall commence our illustration.
\paragraph{The $\hat{E}_6$ Singularity: }
For $\hat{E}_6$, we recall from \cite{Benvenuti:2006qr} that
\beq\label{g1-E6}
f = \frac{1 - t^4 + t^8}{1 - t^4 - t^6
  +  t^{10}} = 1 + t^6 + t^8 + 2\,t^{12} + t^{14} + \cO(t^{16})
:=  \sum\limits_{k=0}^\infty a_k t^k \ .
\eeq
Multiplying through by the denominator gives us
\bea\label{g1-E6-recur}
1 - t^4 + t^8 
&=& \sum\limits_{k=0}^\infty a_k t^k 
  - \sum\limits_{k=4}^\infty a_{k-4} t^k 
  - \sum\limits_{k=6}^\infty a_{k-6} t^k
  + \sum\limits_{k=10}^\infty a_{k-10} t^k \\ \nn
&=& 
\sum\limits_{k=0}^9 a_k t^k - \sum\limits_{k=4}^9 a_{k-4} t^k 
- \sum\limits_{k=6}^9 a_{k-6} t^k +
\sum\limits_{k=10}^\infty \left(
a_k - a_{k-4} - a_{k-6} + a_{k-10}
\right) \ .
\eea
Identifying the coefficients of powers of $t$, this readily gives us
the recursion relation:
\beq\label{g1-E6-init}
a_k = a_{k-4} + a_{k-6} - a_{k-10}, \quad k \ge 10 \ .
\eeq
There should be 10 initial conditions for $a_k$, which could be
obtained by matching the 1 as well as the $-t^4$ and $t^8$ terms in
the LHS with the various finite sum pieces in the RHS of
\eref{g1-E6-recur}. 
Alternatively, it is easier to simply read off the first 10
values of $a_k$ in the series expansion in \eref{g1-E6}, giving us
\[
a_{0,6,8} = 1, \mbox{ else, } a_{k < 10} = 0 \ .
\]

Of course, all linear homogeneous difference equations of this kind can be
solved. Upon substitution of the ansatz $a_k = t^k$ for some $t \in
\IC$, one obtains the eigen-equation for $t$ which is simply the 
denominator $1 - t^4 - t^6 +  t^{10}$ in \eref{g1-E6}. This
has 10 roots: $\{\omega_6^{i=0,\ldots,5},\pm i\}$
with double roots at 1 and $-1$. Using the usual trick that for each
multiple root $\lambda$ of order $m$, there are extra roots
$k^{j = 1, \ldots, m-1} \lambda^k$, the solution is reaily found to be
\bean
a_k &=& {\left( -1 \right) }^k\,\left( c(1) + 
     k\,c(2) \right)  + c(3) + 
  k\,c(4) + c(5)\,\cos
  (\frac{k\,\pi }{3}) + \\
&&  \quad c(6)\,\cos (\frac{k\,\pi }{2}) + 
  c(7)\,\cos (\frac{2\,k\,\pi }{3}) + 
  c(8)\,\sin (\frac{k\,\pi }{3}) + 
  c(9)\,\sin (\frac{k\,\pi }{2}) + 
  c(10)\,\sin (\frac{2\,k\,\pi }{3}) \ ,
\eean
with initial constants $c(i)$, $i=1,\ldots,10$. Matching these with 
the 10 initial conditions in \eref{g1-E6-init} gives us the final solution
\bean
a_k &=& \frac{1}{72}
\left[
3\,\left( 1 + {\left( -1 \right) }^k \right) \,\left( 1 +
  k \right)  +  18\,\cos (\frac{k\,\pi }{2}) +
    24\,\left(\cos (\frac{k\,\pi }{3}) + 
    \cos (\frac{2\,k\,\pi }{3})\right) + \right. \\
&& \qquad \left. + 8\,{\sqrt{3}}\, \left(
\sin(\frac{k\,\pi }{3}) - \sin (\frac{2\,k\,\pi }{3}) \right)
\right], \qquad
k = 0, 1 , \ldots
\eean
There is an obvious cyclicity of 12 and $a_{12m} = 1 + m$ for $m \in
\IZ_{\ge 0}$. We will shortly see this in another guise in
\sref{sec:MacEta}.

\paragraph{The $\hat{E}_7$ Singularity: }
For $\hat{E}_7$, we have \cite{Benvenuti:2006qr} that
\[
f = \frac{1 - t^6 + t^{12}}{1 - t^6 - t^8 + t^{14}} =
1 + t^8 + t^{12} + t^{16} + t^{18} + t^{20} + 2\,t^{24} + t^{26} + t^{28} + 
  t^{30} + 2\,t^{32} + \cO(t^{34})
\ ,
\]
giving us the recursion relations
\beq
a_k = a_{k-6} + a_{k-8} - a_{k-14}, \quad k \ge 14, \qquad
a_{0,8,12} = 1,  \mbox{ else, } a_{k < 14} = 0 \ .
\eeq
This can be readily solved using the above methods to be, for
$k = 0, 1, \ldots$,
\bean
a_k &=& \frac{1}{144}\left[
3\,\left( 1 + {\left( -1 \right) }^k \right) \,\left( 1 + k \right)  + 
2\,\cos (\frac{k\,\pi }{2})\,
  \left( 27 + 24\,\cos (\frac{k\,\pi }{6}) + \right. \right.\\
&&  \qquad \left. \left.
    18\,\left(\cos (\frac{k\,\pi }{4}) - \sin (\frac{k\,\pi
    }{4})\right) - 8\,{\sqrt{3}}\,\sin (\frac{k\,\pi }{6})  \right)
\right] \ .
\eean
Again, there is an obvious cyclicity of 24 and $a_{24m} = 1 + m$ for $m \in
\IZ_{\ge 0}$.

\paragraph{The $\hat{E}_8$ Singularity: }
For $\hat{E}_8$, we have that \cite{Benvenuti:2006qr} 
\bean
f = \frac{1 + t^2 - t^6 - t^8 - t^{10} +
  t^{14} + t^{16}} {1 + t^2 - t^6 - t^8 - t^{10} - t^{12} + t^{16} +
  t^{18}} &=&
1 + t^{12} + t^{20} + t^{24} + t^{30} + t^{32} + t^{36} + t^{40} +
t^{42} + \\
&&t^{44} + t^{48} + t^{50} + t^{52} + t^{54} + t^{56} + 2\,t^{60} +
  t^{62} + \cO(t^{64}) \ ,
\eean
giving us the recursion relations
\[
a_k = -a_{k-2} + a_{k-6} + a_{k-8} + a_{k-10}+ a_{k-12}- a_{k-16}-
  a_{k-18} \quad k \ge 18, \qquad
a_{0,12} = 1,  \mbox{ else, } a_{k < 18} = 0 \ .
\]
Again, this can be solved exactly, giving us, for $k = 0,1, \ldots$,
\bean
a_k &=& \frac{1}{1800}
\left[
15\,\left( 1 + {\left( -1 \right) }^k \right) \,\left( 1 + k \right)  
     + 36\,{\sqrt{5\,\left( 5 - 2\,{\sqrt{5}} \right) }}\,
      \left( \sin (\frac{2\,k\,\pi }{5}) - \sin (\frac{3\,k\,\pi }{5})
      \right) + \right. \\
&& \quad \left.
     + 36\,{\sqrt{5\,\left( 5 + 2\,{\sqrt{5}} \right) }}\,
      \left( \sin (\frac{k\,\pi }{5}) - \sin (\frac{4\,k\,\pi }{5})
      \right) + \right. \\
&& \quad \left.
      +10\,\cos (\frac{k\,\pi }{2})\,
      \left( 45 + 
      36\,\left( \cos (\frac{k\,\pi }{10}) + 
      \cos (\frac{3\,k\,\pi }{10}) \right) + 
      60\,\cos (\frac{k\,\pi }{6})  - 
      20\,{\sqrt{3}}\,\sin (\frac{k\,\pi }{6}) \right) 
\right] \ .
\eean
Once more, there is an obvious cyclicity of 60 and $a_{60m} = 1 + m$ for $m \in
\IZ_{\ge 0}$.

\paragraph{The $\hat{A}_n$ Family: }
Now, let us move on to the infinite families.
For, $\hat{A}_{n-1}$, we have that,
letting $f_\infty(t;~\hat{A}_n) = \frac{(1+t^n)}{(1-t^2)(1-t^n)} :=
\sum\limits_{k=0}^\infty a_k t^k$,
\beq\ba{rcl}
1+t^n &=& \sum\limits_{k=0}^\infty a_k t^k -
  \sum\limits_{k=2}^\infty a_{k-2} t^k -
  \sum\limits_{k=n}^\infty a_{k-n} t^k +
  \sum\limits_{k=n+2}^\infty a_{k-n-2} t^k \\
  &=& 
  \left(a_0 + a_1 t + \ldots + a_{n+1} t^{n+1} \right)-
  \left(a_0t^2 + a_1 t^3 + \ldots + a_{n-1} t^{n+1} \right)- \\
  && -
  \left(a_0t^n + a_{1} t^{n+1} \right)+
  \sum\limits_{k=n+2}^\infty \left(a_k - a_{k-2} - a_{k-n} + a_{k-n-2}
  \right) t^k \ .
\ea\eeq
Identifying coefficients of $t$, we have that
\beq\label{recur-an}
a_k = a_{k-2} + a_{k-n} - a_{k-n-2} , \quad k \ge n+2 \ ;
\eeq
we still need $n+2$ initial conditions. One is obvious, $a_0=1$, the
remaining can be obtained by solving for the system of associated
equations above for $a_1,\ldots,a_{n+1}$.

Now, we could solve this recursion equation, which is rather difficult
because of the determination of these initial conditions. However, in
this case, it is far easier to simply observe the pattern and conclude
that
\beq\label{An-an}\ba{ccc}
n = \odd & & a_k = \floor(\frac{k}{n}) 
    + \frac12\left(1 + (-1)^{\bmod(k, n)}\right) \\
n = \even & & a_k = \left(\floor(\frac{k}{n}\right) + \frac12)
   \left(1 + (-1)^{\bmod(k,n)}\right)
\ea\eeq
Again, the cyclicities are apparent: for odd $n$, $a_{k=2 \beta n}=2
\beta$ and for even $n$, $a_{k=2 \beta n} = 4\beta + 1$ 
for $\beta \in \IZ_{\ge 0}$.
We will write these coefficients explicitly later 
using the MacMahon and Dedekind functions in \sref{sec:MacEta}.

\paragraph{The $\hat{D}_n$ Family: }
For the $\hat{D}_{n+2}$ groups, the recursion
relation reads
\beq
f_\infty(t;~\hat{D}_{n+2})=\frac{(1+t^{2n+2})}{(1-t^4)(1-t^{2n})} :=
\sum\limits_{k=0}^\infty a_k t^k,
\qquad
a_{k} = a_{k-4} + a_{k-2n} - a_{k-2n-4}; \quad k \ge 2n+4 \ ,
\eeq
together with $2n+4$ initial conditions.

Once again, it is easier to directly observe the pattern here.
First, we notice that, upon making the substitution $t^2 \to t$, the
Hilbert series becomes quite analogous to the A-series. 
Indeed, in analogy to \eref{someAexp}, we find that the coefficients
for even $n$ come in periodicity of order $n$ and that for the $k$-th
period the even coefficients are $2k+1$ and the odd coefficients are
0. We will use this in writing expressions for the generating function
in \sref{sec:MacEta}.

In summary, we
may observe the pattern of the expansion coefficients as:
\beq
\frac{(1+t^{n+1})}{(1-t^2)(1-t^{n})} := \sum\limits_{k=0}^\infty b_k
t^k \quad \Rightarrow \quad 
b_k = \frac12\left(1 + (-1)^k\right) + 
   \floor\left( \frac1n \bmod(k, 2n)\right) + 
   2\, \floor\left( \frac{k}{2n} \right) \ .
\eeq
Therefore, upon restoring $t \to t^2$, we only have even powers;
whence, for all $k=0,1,\ldots$,
\beq
a_k = \left\{
\ba{lcc}
0, & \quad & k~\odd;\\
\frac12\left(1 + (-1)^{k/2}\right) + 
   \floor\left( \frac1n \bmod(k, 4n)\right) + 
   2\, \floor\left( \frac{k}{4n} \right), & \quad & k~\even.
\ea
\right.
\eeq

%
\subsection{Full Generating Functions: MacMahon and Euler}
\label{sec:MacEta}
Having obtained analytic expressions for the counting of single-trace
GIO's, i.e., the coefficient of the fundamental generating function
$f_\infty$, the Hilbert series,
we can say something further about the plethystic
exponentials.
The expressions for the ADE orbifolds can be represented as infinite
sums. Such sums appear in different counting formulae for integer
partitions under special restrictions.
For example, it is not surprising to find that the multi-trace
generating function for $\IC^2$, 
\beq
g_\infty (t ; \IC^2) = \exp\biggr(\sum_{n=1}^\infty \frac{1}{n}
\left((1-t^n)^{-2}-1\right)\biggr) = 1+2t^2+6t^3+14t^4+33t^5+ 70t^6+\ldots 
\eeq 
generates the sequence of the number of partitions of $n$ objects with
2 colors \cite{Cameron}. 
It would be interesting to find similar results for the ADE series.

Now, we can use an alternative representation for the generating
functions such as \eref{g-nu}. For the exmaple of $\IC^2$, we
recall that 
\[
g( \nu ; t; \IC^2) = \prod_{n=0}^\infty (1 - \nu t^n) ^ {-(n+1)}. 
\]
We note that the coefficients $a_n$ have a linear piece and a
constant piece. This property turns out to be generic for all 2
dimensional singular manifolds. We will therefore define two basic
functions. First, let the generalized MacMahon function be:
\beq
M(\nu ; t) := \prod_{n=1}^\infty (1-\nu t^n)^{-n} \ ;
\eeq
next, let the generalized Dedekind Eta function (in this form it is
actually the generalised Euler function, which differs from the Eta
function by the famous factor of $t^{-1/24}$) 
be defined as:
\beq
\eta(\nu ; t) := \prod_{n=0}^\infty (1-\nu t^n)^{-1} \ .
\eeq
In terms of these functions we can now rewrite 
\[
g( \nu ; t ;~\IC^2) = M(\nu ;~t) \eta (\nu ;~t) \ .
\]
We now wonder if this form of the expression can be done for the ADE
orbifolds due to the fact that the coefficients $a_n$ for the Hilbert
(Molien) series, as we recall from \cite{Benvenuti:2006qr}, are always of a
linear and a constant form, corresponding to the functions $M$ and
$\eta$, respectively. This turns out to be correct.

Let us look, for example, at the generating function for $\IC^2/\IZ_2$.
We find that
\[
g( \nu ; t ; ~\IC^2/\IZ_2 ) = \prod_{n=0}^\infty (1 - \nu t^{2n} ) ^
{-(2n+1)} \ ,
\]
which can be easily rewritten as
\[
g( \nu ; t ; ~\IC^2/\IZ_2 ) = M ( \nu ; t^2 ) ^ 2 \eta ( \nu ; t^2 ) 
\ .
\]
To proceed with the full A-family, we now use the periodic pattern for
the coefficient $a_n$ which
was obtained above in \eref{An-an}, and obtain
the following succint expression for the generating function $g(\nu,~t)$:
\bean
g(\nu ; t ; \IC^2/\IZ_{2k} ) &=& \prod_{j=0}^{k-1} M( \nu t^{2j} ;
t^{2k} )^2 \; \eta (\nu t^{2j} ; t^{2k} ) \cr 
g(\nu ; t ; \IC^2/\IZ_{2k+1} ) &=& \prod_{j=0}^{2k} M( \nu t^{j} ;
t^{2k+1} ) \; \prod_{j=0}^k \eta (\nu t^{2j} ; t^{2k+1} ) 
\eean

Similarly, we can obtain the full-generating function for the D-family:
\bean
g(\nu, t; \IC^2/\hat{D}_{2k}) &=& \prod_{j=0}^{2k-3}M(\nu t^{2  j} ; t^{4  k -
  4}) \; \prod_{j = 0}^{k - 2}\eta(\nu t^{4  j}; t^{4  k - 4}) \\
g(\nu, t; \IC^2/\hat{D}_{2k+1}) &=& \prod_{j = 0}^{4  k - 3} M (\nu t^{2 j} ;
t^{8  k - 4} )^2\prod_{j = 0}^{2  k - 2}\eta(\nu t^{4  j} ; t^{8  k -
  4} )\prod_{j = 0}^{2  k - 2}\eta ( \nu t^{2  j + 4  k - 2} ; t^{8  k
  - 4} ) \ .
\eean

Finally, for the E-family, as mentioned above
we find that each of the Hilbert series
come with a quasi-periodicity of 12, 24, and 60 for $\hat{E}_{6,7,8}$, 
respectively, which can be seen from the explicit expressions for the
coefficients in the various equations for $a_k$ given in the previous
subsection. 
The growth of the coefficients is always linear in these periods.
Furthermore, odd powers never appear.
Therefore, one can write vectors of length 6, 12, and 30, 
which will denote the starting powers of the coefficients. Explicitly,
we have:
\[\ba{rcl}
v_{E6} &=& \{1,0,0,1,1,0\}\\
v_{E7} &=& \{1,0,0,0,1,0,1,0,1,1,1,0\}\\
v_{E8} &=&
\{1,0,0,0,0,0,1,0,0,0,1,0,1,0,0,1,1,0,1,0,1,1,1,0,1,1,1,1,1,0\} \ .
\ea\]
The generating functions then take the form
\bean
g(\nu, t; \IC^2/\hat{E}_{6}) &=& 
\prod_{j=0}^{5}M(\nu t^{2  j} ; t^{12}) \eta(\nu t^{2  j};
t^{12})^{v_{E_6}^j} \\
g(\nu, t; \IC^2/\hat{E}_{7}) &=& 
\prod_{j=0}^{11}M(\nu t^{2  j} ; t^{24}) \eta(\nu t^{2  j};
t^{24})^{v_{E_7}^j} \\
g(\nu, t; \IC^2/\hat{E}_{8}) &=& 
\prod_{j=0}^{29}M(\nu t^{2  j} ; t^{60}) \eta(\nu t^{2  j};
t^{60})^{v_{E_8}^j} \ .
\eean
Two curiosities are perhaps worthy of note. First, the periodicities
of the coefficients in the Hilbert series are, respectively, one half
the order of the finite groups themselves. Second, for each of the
vectors $v_{E6,E7,E8}$ above, one can draw a line at the middle, then
upon mirror reflection about this line, a zero is mapped to a one, and
vice versa.

%
\subsection{Asymptotic Expansions for $g_\infty$}
As was emphasised in \cite{Benvenuti:2006qr} as well as the proceeding
discussions, the asymptotic behaviour of $g_\infty$ is of great
interest.
Using the Meinardus Theorem, we can estimate the asymptotic behaviour for
$g_\infty$ for the ADE-singularities.
Though the expressions for the $a_k$ are, evidently, quite involved,
the large $k$ behaviour is dominated by the term proportional to $k$,
which can be directly observed;
other eigenvalues have less than unit modulus and decay {\it ad nullam}.
We wish to find $d_m$ in
\[
\prod_{k=1}^{\infty} \frac{1}{(1 - \, t^k)^{a_k}} :=
\sum_{m=0}^\infty d_m t^m
\]
for large $m$.

It suffices to see the large $k$ behaviour of $a_k$ for the
ADE-orbifolds. We see, from the expressions above, that all $a_k$ are
essentially linear in $k$. For $\hat{A}_{n-1}$, $n$ odd, the
coefficient of the linearity is simply $1/n$. For all other cases, the
coefficient is the reciprocal of $1/2$ the order of the
group. However, for all these cases, exactly $1/2$ of the terms are zero
and contribute 1 to the product. Therefore, overall, the effective
large $k$-behaviour is still simply the reciprocal of the order of the
group. Hence, we conclude that
\beq
\mbox{For }G = ADE, \qquad a_k \sim \frac{k}{|G|} \ . 
\eeq

Now, we are at liberty to use the Meinardus analysis. For $a_k \sim
k$, we recall from \cite{Benvenuti:2006qr} that this is the case of
the MacMahon function, whose behaviour goes as
\beq\label{mac-asym}
\prod_{k=1}^{\infty} \frac{1}{(1 - \, t^k)^{k}} :=
\sum_{m=0}^\infty \varphi(m) t^m, \quad \Rightarrow \quad
\varphi(m) \sim
\frac{{2^{-\frac{11}{36}}{\zeta(3)}^{\frac{7}{36}}}e^{\frac{1}{12}}}
{{G_l}\,{\sqrt{3\,\pi }}} \, m^{-\frac{25}{36}} \,
\exp\left( \frac32 
          {\left( 2\,\zeta(3) \right) }^{\frac{1}{3}}
	  m^{\frac{2}{3}}
\right),
\eeq
with $G_l:= \frac{1}{12} - \zeta '(-1)$ being the Glaisha constant.
Thus, we see that for $G$ being an ADE-group,
\beq
d_m \sim \varphi(m)^{\frac{1}{|G|}} \ .
\eeq
In fact, taking the logarithm of this expression will give us the
entropy of the quiver gauge theory as discussed in \sref{s:entropy}. 
We conclude that the entropy is
reduced by a factor of $|G|$ and this is the natural expectation from an
extensive parameter like the entropy since, by the orbifold action, we are
losing $|G|$ of the degrees of freedom.  

Incidentally, the MacMahon function is the generating function for the
plane-partition problem which is a generalisation of the
Young Tableaux to 2-dimensions. That is, consider an integer $m$, how
many ways are there to write
\[
m = \sum_{i,j} n_{i,j} \quad \mbox{such that} \quad
n_{i+1,j} \ge n_{i,j},~n_{i,j+1} \ge n_{i,j};\quad n_{i,j} \in \IZ_{+}
\ .
\]
The answer was shown in \cite{macmahon} to be precisely $\varphi(m)$.
The 1-dimensional partition problem, i.e., how many Young Tableaux
(also called Ferrers Diagram) are there of a given total number of
squares, is simply the standard partitioning problem. By this we mean
how many ways, irrespectively ordering, are there to write a given
integer $m$ as sums of integers. This is because we could
always order the parts in decreasing fashion and arrive at a Young
Tableau. The generating function here is simply the famous Euler
function $\prod\limits_{k=1}^{\infty} \frac{1}{(1 - \, t^k)}$.
It is a curious fact that 3 and higher dimensional analogues of the
problem remain unsolved. A conjecture was made in \cite{macmahon}
which was later shown to be incorrect.

We see that counting GIO's for the ADE gauge theories is related to
the 2-dimensional counting problem in a simple fashion: generating
function, asymptotically, 
is simply that of the MacMahon to the $|G|$-th root. This can be
conceived of tiling, asymptotically, 
not the whole plane, but rather, a $|G|$-th
fraction of the plane, as the orbifold indeed requires. However, to 
which exact partition problems the ADE results correspond remains
elusive.

%
\subsection{The MacMahon Conjecture}\label{s:macmahon}
One could imagine what the result for solid-partitions,
which, as mentioned above, is unknown, might actually be.
Let us tabulate the result for the gauge theories for $\IC$ and
$\IC^2$. We recall from \cite{Benvenuti:2006qr} that
\bean
&&f_\infty(t;~\IC) = \frac{1}{1-t} = \sum_{k=0}^\infty t^k, 
\quad \Rightarrow \quad 
g_\infty(t;~\IC) = \prod_{k=1}^{\infty} \frac{1}{(1 - \, t^k)};\\
&&f_\infty(t;~\IC^2) = \frac{1}{(1-t)^2} = \sum_{k=0}^\infty (k+1) t^k, 
\quad \Rightarrow \quad 
g_\infty(t;~\IC^2) = \prod_{k=1}^{\infty} \frac{1}{(1 - \, t^k)^{k+1}}
\ .
\eean
Thus we see that the multi-trace problem for $\IC$ counts the
1-dimensional partition; that for $\IC^2$, when shifted by 1, counts
the 2-dimensional problem. It is perhaps natural to guess that the one
for $\IC^3$, when shifted by one, would give the generating function for
the 3-dimensional partition problem, i.e.,
\bea\label{solid-guess}
&&f_\infty(t;~\IC^3) = \frac{1}{(1-t)^3} = \sum_{k=0}^\infty
\frac{(k+1)(k+2)}{2} t^k, \quad \mbox{shift} \Rightarrow a_k =
\frac{k(k+1)}{2}  \quad \Rightarrow
\\ \nn
&&g_\infty = \prod_{k=1}^{\infty} \frac{1}{(1 - \,
  t^k)^{\frac{k(k+1)}{2}}}  
= 1 + t + 4\,t^2 + 10\,t^3 + 26\,t^4 + 59\,t^5 + 141\,t^6 + 
  310\,t^7 + 692\,t^8 + \cO(t^{9}) \ .
\eea
Unfortunately, this leads us back to MacMahon's erroneous guess
\cite{macmahon}. The correct numbers, as generated by exhaustive computer
simulation of the explicit partitions, should be
(cf.~e.g.~\cite{sloane}):
\beq\label{true-solid}
1, 1, 4, 10, 26, 59, 140, 307, 684, 1464, 3122, 6500, 13426, 27248,
54804, 108802,  \ldots
\eeq
One sees that starting from the term 141, the generating function
$g_\infty$ in \eref{solid-guess} over-counts. The actual series $a_k$
which {\it does} generate the correct numbers can be easily found, by
taking the plethystic logarithm, to be (cf.~also \cite{sloane})
\beq\label{true-ak-solid}
\ba{rcl}
a_{k=1,2,\ldots} &=& \{
1, 3, 6, 10, 15, 20, 26, 34, 46, 68, 97, 120, 112, 23, -186, -496,\\
&&-735, -531, 779, 3894, 9323, 16472, 23056, 23850, 10116, \ldots
\} \ .
\ea\eeq
As it is evident, the negative entries complicate things; it suggests
that $a_k$ itself cannot be a Hilbert series. Could it be the
plethystic logarithm of a Hilbert series? Recall that for these,
there are often negative entries, signifying relations among
fundamental invariants. Well, it certainly is the plethystic logarithm
of something; this is after all, how the series \eref{true-ak-solid}
was obtained from \eref{true-solid}, but this brings us to to where we
started.
What if we took the plethystic
log of \eref{true-ak-solid} itself? Unfortunately, we obtain nothing
particularly enlightening.

%
%
\section{All $SU(3)$ Subgroups}\label{s:su3}\setequation
We have discussed the ADE-groups above in some detail; of perhaps more
physical interest are the orbifolds of $\IC^3$. These are local
Calabi-Yau threefolds that give rise to $\cN=1$ 4-dimensional chiral
gauge theories on the D3-brane world-volume.
The quiver theories were studied in \cite{Hanany:1998sd} and using the
notation therein, the discrete finite
subgroups of $SU(3)$ are:
\begin{itemize}
\item[(I)] The infinite family $\IZ_m \times \IZ_n$;
\item[(II)] The infinite families $\Delta(3n^2)$ and $\Delta(6n^2)$;
\item[(III)] The exceptionals $\Sigma_{60}$,
  $\Sigma_{108}$, $\Sigma_{168}$,
  $\Sigma_{216}$, $\Sigma_{648}$, and $\Sigma_{1080}$.
\end{itemize}
The theory of Molien series and algebraic
invariants is nicely exposed in \cite{yauyu}, wherein some explicit
Molien series are also computed for the discrete subgroups of
$SL(3;\IC)$.

In order to explicitly write the generators of the groups, first, define
\[
\omega_n := \exp(\frac{2 \pi i}{n}),
\]
and the matrices
\beq\label{genSU3}
\ba{ccc}
S := \tmat{ 1 & 0 & 0 \cr 0 & \omega_3 & 0 \cr 0 & 0 & {\omega_3}^2
  \cr  },
&
S_1 := \tmat{ 1 & 0 & 0 \cr 0 & {\omega_5}^4 & 0 \cr 0 & 0 & \omega_5
  \cr},
&
S_2 := \tmat{ \omega_7 & 0 & 0 \cr 0 & {\omega_7}^2 & 0 \cr 0 & 0 &
  {\omega_7}^4 \cr};
\\&&\\
T := \tmat{ 0 & 1 & 0 \cr 0 & 0 & 1 \cr 1 & 0 & 0 \cr  },
&
T_1 := 
\frac{1}{\sqrt{5}}\tmat{1 & 1 & 1 \cr 
  2 & \frac12(-1-\sqrt{5}) & \frac12(-1+\sqrt{5}) \cr
  2 & \frac12(-1+\sqrt{5}) & \frac12(-1-\sqrt{5})},
&
T_2 := \smat{ 1 & 0 & 0 \cr 0 & 0 & 1 \cr 0 & -1 & 0 \cr  };
\\&&\\
&
R := -\frac{1}{\sqrt{-7}} 
  \tmat{ -{\omega_7}^3 + {\omega_7}^4 & {\omega_7}^2 - {\omega_7}^5 &
    \omega_7 - 
  {\omega_7}^6 \cr {\omega_7}^2 - {\omega_7}^5 & \omega_7 -
  {\omega_7}^6 & -{\omega_7}^3 +   
   {\omega_7}^4 \cr \omega_7 - {\omega_7}^6 & -{\omega_7}^3 +
   {\omega_7}^4 & {\omega_7}^2 -
   {\omega_7}^5 \cr  };
&
\\&&\\
U :=
\tmat{ {\omega_9}^4 & 0 & 0 \cr 0 & {\omega_9}^4 & 0 \cr 0 & 0 &
  \omega_3\,{\omega_9}^4 \cr },
&
U_1 := \tmat{ -1 & 0 & 0 \cr 0 & 0 & -1 \cr 0 & -1 & 0 \cr  };
&
\\&&\\
V :=
\frac{1}{\sqrt{-3}}\tmat{ 1 & 1 & 1 \cr 1 & \omega_3 & {\omega_3}^2 \cr 1 &
  {\omega_3}^2 & \omega_3 \cr  },
&
V_1 := \frac{1}{\sqrt{5}}
  \tmat{1 & \frac14(-1+\sqrt{-15}) & \frac14(-1+\sqrt{-15}) \cr
  \frac12(-1-\sqrt{-15}) & \frac12(-1-\sqrt{5}) & \frac12(-1+\sqrt{5}) \cr
  \frac12(-1-\sqrt{-15}) & \frac12(-1+\sqrt{5}) & \frac12(-1-\sqrt{5})
  \cr};
&\\&&\\
A(m) := \tmat{\omega_m & 0 & 0 \cr 0 & 1 & 0 \cr 0 & 0 & \omega_m^{-1}},
& 
B(n) := \tmat{1&0&0 \cr 0 & \omega_n & 0 \cr 0 & 0 & \omega_n^{-1}} \ .
&
\\
\ea
\eeq
Finally, we adhere to the usual notation that
\[
G = \gen{g_1, \ldots, g_k}
\]
is the finite group $G$ generated by matrices $g_1,\ldots g_k$.

\subsection{The Abelian Series: $\IZ_m \times \IZ_n$}
The first of our series is simply $\IZ_m \times \IZ_n =
\gen{A(m),~B(n)}$. The Molien series is given by
\bea
&& f(t;~\IZ_m \times \IZ_n) 
= \frac{1}{mn} \sum_{i=0}^{m-1}
\sum_{j=0}^{n-1} \det\left(
\II_{3 \times 3} - t \tmat{\omega_m^i &0&0 \cr
0&\omega_n^j&0\cr 0&0&\omega_m^{-i}\omega_n^{-j}
}
\right)^{-1} \\ \nn
&=&
{1\over m n}\sum_{i=0}^{m-1} \sum_{j=0}^{n-1}{1\over (1-t
\omega_{m}^{i})(1-t\omega_{n}^j) (1-t \omega_{m}^{-i}
\omega_{n}^{-j})}
=
\frac{1}{mn} \sum_{i=0}^{m-1}\sum_{j=0}^{n-1}
\sum_{p,q,r=0}^\infty t^p \omega_m^{ip} t^q \omega_n^{jq} 
  t^r \omega_m^{-ir}\omega_n^{-jr} \ .
\eea
Using the identity
\[
\sum_{i=0}^{m-1} \omega_m^{i x} = m \delta_{x, m\IZ} \ , 
\]
where the Kronecker-Delta is 1 whenever $x$ is a multiple of $n$,
we can see that non-zero contributions come from
\[ 
p= r+ \W{p}\ m~ \mbox{ for } \W p=-[{r\over m}],-[{r\over m}]+1,\ldots; \qquad
q= r+ \W{q}\ n~ \mbox{ for } \W q=-[{r\over n}],-[{r\over n}]+1, \ldots .
\]
Here, $[{r\over m}]$ means to take the integer part (i.e.,
\mbox{floor}$(r/m)$). Whence,
\[
 f(t;~\IZ_m \times \IZ_n) =
\sum_{r=0}^{\infty} \sum_{\W p=-[{r\over m}]}^{\infty}
\sum_{\W q=-[{r\over n}]}^\infty t^{3r+ \W p \ m + \W q \ n} =
\sum_{r=0}^{\infty} { t^{3r-[{r\over m}] m-[{r\over n}]n} \over
(1-t^{m}) (1-t^{n})} \ .
\]
To go further, we can write $r= \W r+ LCM(m,n) z$
for $z=0,1,..., \infty$ and $\W
r=0,1,...,LCM(m,n)-1$, where $LCM$ is the lowest common multiple.
Using this parametrization, the sum reduces to
\bea\label{f-zmzn}
f(t;~\IZ_m \times \IZ_n) &=& 
{1\over (1-t^{m}) (1-t^{n})} \sum_{\W r=0}^{LCM(m,n)-1} 
\sum_{z=0}^\infty t^{3
\W r-[{\W r\over m}] m-[{\W r\over n}]n} t^{LCM(m,n) z} \\ \nn
&=& {1\over (1-t^{m})
(1-t^{n})(1-t^{LCM(m,n)})} \sum_{\W r=0}^{LCM(m,n)-1}  t^{3 \W r-[{\W
      r\over m}] m-[{\W r\over n}]n}\ .
\eea
The above expression becomes particularly simple in the case of
perhaps greatest interest, viz, when $m=n$; here $LCM(m,n)=m$ and 
within the range of summation of $\W{r}$, $[{\W r\over m}]$ is
zero, hence
\beq\label{f-zmzm}
f(t;~\IZ_m \times \IZ_m) = \frac{1-t^{3m}}{(1-t^3)(1-t^m)^3} \ .
\eeq
Taking the plethystic logarithm of \eref{f-zmzm} gives polynomials,
suggesting that $\IC^3/(\IZ_m \times \IZ_m)$ are all complete
intersections! Explicitly, we have that
\beq\label{f1-zmzm}
f_1(t;~\IZ_m \times \IZ_m)  = PE^{-1}[f(t;~\IZ_m \times \IZ_m)]
= \left\{
\ba{ccc}
3t && m=1,\\
3t^2 + t^3 - t^6 && m=2,\\
4t^3 - t^9 && m=3,\\
t^3 + 3t^m - t^{3m} && m \ge 4.
\ea
\right.
\eeq
The $m=1$ case is a good check; this is simply the result for the
parent $\IC^3$ theory.

\paragraph{Refinement: }
As was pointed out in \cite{Benvenuti:2006qr}, where there are enough
isometries, as certainly is the case with toric varieties, refinements
can be made to the Molien series. The above Abelian series are indeed
toric, hence one could write the refined Molien (Hilbert) series as
\[
f(t_1,t_2,t_3;~\IZ_m \times \IZ_n) =
\frac{1}{mn} \sum_{i=0}^{m-1}
\sum_{j=0}^{n-1} \det\left(
\II_{3 \times 3} - \tmat{t_1&0&0\cr 0&t_2&0\cr 0&0&t_3} 
\tmat{\omega_m^i &0&0 \cr
0&\omega_n^j&0\cr 0&0&\omega_m^{-i}\omega_n^{-j}
}
\right)^{-1} \ .
\]
This can, using the above reparametrisation of the summation
variables, be re-written as
\[
f(t_1,t_2,t_3;~\IZ_m \times \IZ_n) =
{1\over (1-t_1^{m})
(1-t_2^{n})(1-t_3^{LCM(m,n)})} 
\sum_{\W{r}=0}^{LCM(m,n)-1}  
(t_1t_2t_3)^{\W{r}}
t_1^{-[{\W{r} \over m}] m} t_2^{-[{\W r\over n}]n} \ .
\]
Once again, for the case of $m=n$, the expression simplifies
considerably:
\beq\label{zmzmf}
f(t_1,t_2,t_3;~\IZ_m \times \IZ_m) =
\frac{1-(t_1t_2t_3)^{m}}{(1-t_1t_2t_3)(1-t_1^m)(1-t_2^m)(1-t_3^m)} \ .
\eeq
The plethystic logarithm of this expression becomes particularly simple:
\beq\label{zmzmf1}
f_1(t_1,t_2,t_3;~\IZ_m \times \IZ_m) = 
t_1^m + t_2^m + t_3^m + t_1 t_2  t_3 - ( t_1 t_2 t_3 )^m \ , 
\qquad m =1,2,3,\ldots
\eeq

\comment{
\paragraph{The $\IC^3 / \IZ_n$ Example}
For completeness, let us include an anlysis for the case of the
Abelian orbifold $\IC^3/\IZ_n$ with action $(a,b,-a-b)$. The Hilbert
series is given by
\bean 
f(t;~\IC^3/\IZ_n) & = & 
{1\over n}\sum_{k=0}^{n-1} {1\over (1-t \omega_n^{ak})(1-t
\omega_n^{bk})(1-t \omega_n^{(-a-b)k})} \\ & = & {1\over n}\sum_{k=0}^{n-1}
\sum_{p,q,x} t^{p+q+x} \omega_n^{k (ap+bq-(a+b)x)} = \sum_{p,q,x}
t^{p+q+x} \delta( (ap+bq-(a+b)x)) 
\eean
It is difficult to proceed further for arbitrary $a,b$, thus assume
action is not degenerate and set $a=1$ and $b$ arbitrary. 
Substituting $a=1$ we have $f(t) = \sum_{q,x} t^{x(b+2)+q(1-b)}$.
However, the delta-function requires that
$x(1+b)\geq b q$ for $q\leq x+{b\over x}$.
Write $x= b \W x+ s$ with $s=0,...,b-1$, we have $q\leq (b+1)\W x+s$;
hence 
\bean 
f(t;~\IC^3/\IZ_n^{(1,b,-1-b)}) 
&=& \sum_{s=0}^{b-1} \sum_{\W x=0}^{\infty}\sum_{q=0}^{(b+1)\W
x+s} t^{q(1-b)+ b(b+2) \W x+ (b+2) s}
= \sum_{s=0}^{b-1} \sum_{\W x=0}^{\infty} { t^{b(b+2)\W x+ (b+2)
s}-t ^{(2b+1) \W x+s +1-b}\over 1-t^{1-b}}\\ & = & {1\over
1-t^{1-b}} \left( {1\over 1-t^{b+2}}-{t^{1-b}(1-t^{b})\over
(1-t^{2b+1})(1-t)}\right)
\eean
}

\subsection{Non-Abelian Subgroups}
Having expounded upon the $\IZ_m \times \IZ_n$ series in detail, we
can proceed to the non-Abelian groups. The Molien series for the
exceptionals can be quite
simply computed by \cite{gap} and are presented in the next
subsection. The two Delta-series maybe dealt with in much the same
manner as the abovementioned $\IZ_m \times \IZ_m$.

The elements of $\Delta(3n^2) := \gen{A(n), B(n), T}$ 
fall into three classes, viz, the
orbits of $\IZ_n^2 \simeq \gen{A(n),B(n)}$ 
under $\{\II, T, T^2\}$ since the matrix
$T$, which we recall from \eref{genSU3}, is of order 3. Therefore,
\bea \nn
f(t;~\Delta(3n^2)) &=& \frac{1}{3n^2}
[
\sum_{i,j=0}^{n-1}  
\det\left(
\II_{3 \times 3} - t \tmat{\omega_n^i &0&0 \cr
0&\omega_n^j&0\cr 0&0&\omega_n^{-i-j}}\right)^{-1}
+
\det\left(
\II_{3 \times 3} - t \tmat{0&\omega_n^j&0\cr
  0&0&\omega_n^{-i-j}\cr\omega_n^i &0&0 \cr}\right)^{-1}\\ \nn
&&
+
\det\left(
\II_{3 \times 3} - t \tmat{0&0&\omega_n^{-i-j}\cr
\omega_n^i &0&0 \cr 0&\omega_n^j&0}\right)^{-1}
]\\ \nn
&=& \frac{1}{3n^2}\left[
n^2\frac{1-t^{3n}}{(1-t^3)(1-t^n)^3}+
n^2\frac{1}{1-t^3}+n^2\frac{1}{1-t^3}
\right]\\  
\label{f-delta3}
&=&\frac{1-t^n+t^{2n}}{(1-t^3)(1-t^n)^2} \ .
\eea
One could in fact take the plethystic logarithm and see that these are
complete intersections:
\beq\label{f1-delta3}
f_1(t;~\Delta(3n^2))  = PE^{-1}[f(t;~\Delta(3n^2))]
= \left\{
\ba{ccc}
t + t^2 + 2\, t^3 - t^6 && n=1,\\
t^2 + t^3 + t^4 + t^6 - t^{12} && n=2,\\
2\, t^3 + t^6 + t^9 - t^{18} && n=3,\\
t^3 + t^n + t^{2n} + t^{3n} - t^{6n}&& n \ge 4.
\ea
\right.
\eeq

In complete analogy, $\Delta(6n^2) := \gen{A(n), B(n), T, T_2}$.
In fact $\Delta(6n^2) \simeq \Delta(6(2n)^2)$, thus it suffices to
consider only odd $n$, and we have that
\[
f(t;~\Delta(6n^2)) = \frac{1+t^{6n+3}}{(1-t^6)(1-t^{2n})(1-t^{4n})},
\qquad n = 1,3,5, \ldots \ .
\]
Again, taking the plethystic logarithm shows these to be complete
intersections 
\beq\label{f1-delta6}
f_1(t;~\Delta(6n^2))  = PE^{-1}[f(t;~\Delta(6n^2))]
= \left\{
\ba{ccc}
t^2 + t^4 + t^6 + t^9 - t^{18}&& n=1,\\
2\, t^6 + t^{12} + t^{21} - t^{42}&& n=3,\\
t^6 + t^{2n} + t^{4n} + t^{6n+3} - t^{12n+6}&& n \ge 5.
\ea
\right.
\eeq

%
\subsection{Summary of $SU(3)$ Subgroups}
We now summarise and tabulate the relevant results for all the
discrete subgroups of $SU(3)$. The Molien series for the exceptional
ones have been computed in \cite{yauyu}. The above results for the infinite
families are new. In addition, we compute the plethystic logarithm of
the Molien series, which should give us the defining equations; a
remarkable fact is that all of them (except $\IZ_m \times \IZ_n$ for
$m \ne n$, whose Molien series we have not been able to simplify
further) are complete intersections.
We use the notation (cf.~\cite{Gray:2006jb} for gauge theory
moduli spaces of this type)
\[
(p,(d_1,\ldots,d_k);~(\ell_1^{a_1},\ldots,\ell_q^{a_q}))
\]
to denote the intersection of $k$ equations, of degrees
$d_1,\ldots,d_k$ in $\IC^p$, composed of $a_1$ invariants of degree
$\ell_1$, $a_2$ invariants of degree $\ell_2$, etc.

The generating functions (Molien series) $f = f_\infty(t)$ 
and the associated $f_1$ which encode the syzygies (defining
equations) for the discrete, finite subgroups of $SU(3)$ are:
\beq{\hspace{-1in}
\ba{|c|c|c|c|c|} \hline
G \subset SU(3) & \mbox{Generators} & \mbox{Molien } f(t; G) 
& f_1 = PE^{-1}(f) & \mbox{Defining Equation} \\ \hline
\IZ_m \times \IZ_n & \gen{A(m),~B(n)} & \mbox{q.v.~\eref{f-zmzn}} & 
\mbox{q.v.~\eref{f1-zmzm}}
& - \\ \hline \hline
\Delta(3n^2) & \gen{A(n),~B(n),~T} & 
\frac{1-t^n+t^{2n}}{(1-t^3)(1-t^n)^3}
& \mbox{q.v.~\eref{f1-delta3}}& \mbox{q.v.~\eref{eq-delta3}} \\ \hline
\Delta(6n^2) & \gen{A(n),~B(n),~T,~T_2} n\mbox{ odd}& 
\frac{1+t^{6n+3}}{(1-t^6)(1-t^{2n})(1-t^{4n})}
& \mbox{q.v.~\eref{f1-delta6}}& \mbox{q.v.~\eref{eq-delta6}} \\ \hline \hline
\Sigma_{60} & \gen{S_1,~T_1,~U_1} & 
  \frac{1 + t^{15}}{\left( 1 - t^2 \right) \,\left( 1 - t^6 \right)
    \,\left( 1 - t^{10} \right) }
& t^2 + t^6 + t^{10} + t^{15} - t^{30} & (4,(30);~(2,6,10,15)) \\ \hline
\Sigma_{108} & \gen{S,~T,~V} &
  \frac{1 + t^9 + t^{12} + t^{21}}{{\left( 1 - t^6 \right) }^2\,\left(
    1 - t^{12} \right) } 
& 2\,t^6 + t^9 + 2\,t^{12} - t^{18} - t^{24} & (5,(18,24);~(6^2,9,12^2))
\\ \hline
\Sigma_{168} & \gen{S_2,~T,~R} & 
  \frac{1 + t^{21}}{\left( 1 - t^4 \right) \,\left( 1 - t^6 \right)
    \,\left( 1 - t^{14} \right) } & 
  t^4 + t^6 + t^{14} + t^{21} - t^{42} & (4,(42);~(4,6,14,21))
\\ \hline
\Sigma_{216} & \gen{S,~T,~V,~UVU^{-1}} &
  \frac{1 + t^{12} + t^{24}}{\left( 1 - t^6 \right) \,\left( 1 - t^9
    \right) \,\left( 1 - t^{12} \right) } & 
  t^6 + t^9 + 2\,t^{12} - t^{36} & (4,(36);~(6,9,12^2))
\\ \hline
\Sigma_{648} & \gen{S,~T,~V,~U} & 
  \frac{1 + t^{18} + t^{36}}{\left( 1 - t^9 \right) \,\left( 1 -
    t^{12} \right) \,\left( 1 - t^{18} \right) } & 
  t^9 + t^{12} + 2\,t^{18} - t^{54} & (4,(54);~(9,12,18^2))
\\ \hline
\Sigma_{1080} & \gen{S_1,~T_1,~U_1,~V_1} & 
  \frac{1 + t^{45}}{\left( 1 - t^6 \right) \,\left( 1 - t^{12} \right)
    \,\left( 1 - t^{30} \right) } & 
  t^6 + t^{12} + t^{30} + t^{45} - t^{90} & (4,(90);~(6,12,30,45))
\\ \hline
\hline \ea}
\eeq
In the table, the defining equations for the Delta-series are:
\beq\label{eq-delta3}
\Delta(3n^2) \simeq
\left\{
\ba{ccc}
(4,(6);~(1,2,3,3)) && n=1,\\
(4,(12);~(2,3,4,6)) && n=2,\\
(4,(18);~(3,3,6,9)) && n=3,\\
(4,(6n);~(3,n,2n,3n)) && n \ge 4.
\ea
\right.
\eeq
and
\beq\label{eq-delta6}
\Delta(6n^2)_{n\mbox{ odd}} \simeq
\left\{
\ba{ccc}
(4,(18);~(2,4,6,9)) && n=1,\\
(4,(42);~(6,6,12,21)) && n=3,\\
(4,(12n+6);~(6,2n,4n,6n+3)) && n \ge 5.
\ea
\right.
\eeq
The exceptional groups are addressed in \cite{yauyu} and our
defining equations, obtained from $f_1$, agrees completely with Theorem
C of p7 therein. The forms of the actual equations, with the
coefficients, are very complicated and the reader is referred to the
aforementioned theorem in {\it cit. ibid.}

\subsection{The Fundamental Generating Function: The Hilbert Series}
\label{s:hilbert}
Before we proceed to discuss other fascinating features of
$\IC^3$-orbifolds in the ensuing section,
let us venture on a small digression. In many expressions above, we
have seen the power of the plethystic programme: how the plethystic
logarithm of the Molien series encodes the geometrical information of
the orbifold, with the situation even more conspicuous for complete
intersections. We advertised in the introduction and in
\cite{Benvenuti:2006qr}, the paramountcy of the fundamental generating
function $f_\infty = g_1$, here we shall explain why it should capture
the geometry.

Let us give the formal definition of the Hilbert Series
(cf.~e.g.~\cite{eisenbud}). Let $M := \bigoplus\limits_{i}M_i$ 
be a graded module over 
$K[x_1, \ldots ,x_n]$ (for $K$ some field)
with respect to weights $w_1, \ldots w_n$,
then the Hilbert Series is the generating function for the dimension
of the graded pieces:
\[
H(t; M) := \sum\limits_i \dim_K(M_i) t^i \ .
\]
Usually, we take $K$ to be $\IC$ and are working over polynomials in
$n$ variables; in this case, the grading $i$ can be taken to be the
total degree and $\dim_K(M_i)$ is simply the number of independent
polynomials at degree $i$. The fundamental property of the Hilbert
Series is that it is, in fact, a rational function, of the form
\beq\label{HS}
H(t; M) = \frac{Q(t)}{\prod\limits_{i=1}^n (1-t^{w_i})} \ ,
\eeq
where $Q(t)$ is some, in general rather complicated, polynomial.

In the case of orbifolds, the Molien series counts the invariant
polynomials of a given degree. Since the syzygies (relations) 
of these invariants define the orbifold as a variety, the Molien
series is therefore the Hilbert series for the orbifold
\cite{yauyu}. It just so happens that in this case, we have a nice way
to compute the Hilbert series, using the data of the finite group,
viz., expression \eref{molien}. In the case of toric singularities,
the situation is similar, the equivariant index of
\cite{Martelli:2006yb} and the equivalent sum over vertices in
$(p,q)$-webs in \cite{Benvenuti:2006qr}, reduces the evaluation of the
Hilbert series to combinatorics of the toric diagram.

Let us illustrate the foregoing generalities. Take our familiar
$\Delta(27)$ orbifold; we recall from \eref{f-delta3} that
$f_\infty(t;\Delta(27) = \frac{1-t^3+t^{6}}{(1-t^3)^3}$. Now, this can
in fact be re-written into what was dubbed ``Euler form'' in
\cite{Benvenuti:2006qr}, i.e.,
\beq\label{g1-delta27}
f_\infty(t;\Delta(27) = \frac{1-t^3+t^{6}}{(1-t^3)^3} =
\frac{1 - t^{18}}
  {{\left( 1 - t^3 \right) }^2\,
    \left( 1 - t^6 \right) \,\left( 1 - t^9 \right) } \ .
\eeq
In this form, both numerator and denominator are products of
$(1-t^{w_i})$ factors. Then, from \eref{HS}, geometrically,
$\IC^3/\Delta(27)$ could be realised in $\IC[x_1, \ldots, x_4]$ with
weights $(3,3,6,9)$. This choice of weights arises because the 4
primitive invariants polynomials in the coordinates $(x,y,z)$ of
$\IC^3$ are respectively of degrees 3,3,6 and 9.
Recall further, from \S 5.2 of
\cite{Benvenuti:2006qr}, that ($m_i, w_j \ge 0$ and not necessarily
distinct)
\[
PE^{-1}\left[
\frac{\prod\limits_i (1-t^{m_i})}{\prod\limits_j (1-t^{w_j})}
\right] = 
\sum_i t^{m_j} - \sum_j t^{w_j} \ ,
\]
we have from \eref{g1-delta27} that
\[
f_1(t; \Delta(27)) = PE^{-1}[f_\infty(t; \Delta(27))] =
3\, t^3 + t^6 + t^9 - t^{18} \ ,
\]
in agreement with \eref{f1-delta3}.
Therefore, indeed the Hilbert series has the promised properties and
indeed we see why $f_1$ should encode the geometric information of the
variety.

Two cautionary notes. Though the denominator of the Hilbert series is
always in Euler form, specifiying essentially the information about
the embedding space, the numerator $Q(t)$ is in general
complicated. When $Q(t)$ can indeed be placed into Euler form, $f_1$
terminates and $\cM$ is a complete intersection; otherwise, $f_1$ is
an infinite series, encoding progressively higher syzygies. Second,
the form of the Hilbert series is sensitively dependent on the choice
of embedding. Had we not chosen the weights $(3,3,6,9)$ for the above
example, but, rather, have simply tried to find relations among the 4
primitive invariants, we would have found a complete intersection
whose Hilbert series is $\frac{(1-t^{18})}{(1-t)^4}$, which would not
have given enough information about the geometry of the orbifold.

%
\section{Discrete Torsion}\label{s:dis}\setequation
One might wonder what happens if one were turn on discrete torsion for
the orbifold probe theories. In
the D-brane probe context, this was initiated by
\cite{Douglas:1998xa,Douglas:1999hq}. 
In \cite{Feng:2000af}, it was realised that the
most systematic approach is to compute the so-called covering group
$\tilde{G}$ of the orbifold group $G$. The discrete torsion then
corresponds to the second group-cohomology $A = H^2(G,U(1))$, which is
an Abelian group (so-called Schur multiplier) such that $\tilde{G}/A
\simeq G$.

For all subgroups of $SU(2)$, the Schur multiplier is trivial and
hence the corresponding $\cN=2$ gauge theories do not admit discrete
torsion. For the subgroups of $SU(3)$, however, the situation is more
interesting and the discrete-torsion and the corresponding Schur
multipliers and covering groups have been computed and classified in 
\cite{Feng:2000af}.

The moduli space
for the discrete torsion theories for $\IZ_n \times \IZ_n$ has been
expounded in detail in \cite{Douglas:1999hq} (cf.~also
\cite{Berenstein:2000ux} for a non-commutative perspective). 
In general (cf. Section 3.2 of \cite{Douglas:1999hq}), 
for $N$ D3-branes, it is a $U(N)$ theory with 3
adjoints $\phi_{i=1,2,3}$ which are $N \times N$ matrices, and with a
superpotential
\beq\label{superDis}
W = \tr\left[ \phi_1 (\phi_2 \phi_3 - \omega_n^{-1} \phi_3 \phi_2)
  \right] \ .
\eeq
As an illustration, let us first study the simplest case of $N=1$. Here, 
the superpotential is $W=(1-\omega_n^{-1}) \phi_1 \phi_2 \phi_3$ and
the gauge invariants are simply the 3 numbers $\phi_{i=1,2,3}$.
The moduli space is therefore just the F-flat solutions, which are 
$\phi_1 \phi_2 = \phi_2 \phi_3 = \phi_3 \phi_1 = 0$.
Hence, the moduli space $\cM$ consists of 3 branches, all touching at
the origin: the first parametrized by $\phi_1$ non zero and
$\phi_2=\phi_3=0$, and the other 2 being cyclic permutations.

To construct the generating function we will use a notion
called surgery \cite{HR}. It is trivial for
this case but is generically powerful for more involved cases. Since
each branch of $\cM$ is the complex line we have three $U(1)$
isometries (this holds true for higher $N$ as well) 
and $g_1$ gets a contribution $1/(1-t_i)$ for each $i=1,2,3$.
We sum all together as the spaces are not intersecting at generic
points but need to subtract the intersection spaces which here is just
the one point at the origin. The result for the fundamental generating
function is thus
$g_1 = 1/(1-t_1) + 1/(1-t_2) + 1/(1-t_3) -2$
Setting $t_i=t$ gives $f_\infty(t) = g_1(t) = 3/(1-t)-2$. Taking the
plethystic logarithm gives us an infinite series
$3t - 3t^2 + 2t^3 - \ldots$, whose first two terms
whereby agrees with the F-flat equation above.

\subsection{Example: $\IZ_2 \times \IZ_2$}
Now, how do we reproduce the above quantities using our Hilbert series
and plethystic programme? Let us consider in detail 
the $Z_2\times Z_2$ example. With our discrete torsion,
we have one $U(N)$ gauge group with three chiral fields and
superpotential $W= \tr(XYZ+XZY)$. Thus the F-terms induce 
anti-commutative relations, viz., $XY=-YX, XZ=-ZX, YZ=-ZY$. This is 
the simplest example which allows discrete torsion and 
\cite{Douglas:1999hq} claims that the solution
(F-terms plus D-terms) is given by
\bea 
X= X_1\otimes \sigma_1,~~~Y= Y_1\otimes \sigma_2,~~~Z=
Z_1\otimes \sigma_3,
\eea
where $\sigma_i$ are Pauli matrices and  $X_1, Y_1, Z_1$ all
commute and so can be chosen to be all diagonal.

To construct the single-trace
gauge invariant mesonic operators, we write down the
general form as $\tr(X^{n_1} Y^{n_2} Z^{n_3})$ with
$n_i=0,...,\infty$. Because $\sigma_i^2=I$ we can divide such
operators into eight cases
\begin{itemize}

\item (1) All $n_i$ are even, i.e., $n_i=2k_i$.  In this case we have
  the sum as
\bean 
\sum_{k_i=0}^{\infty}
(t_1^2)^{k_1}(t_2^2)^{k_2}(t_3^2)^{k_3}={1\over (1-t_1^2)  (1-t_2^2)
 (1-t_3^2)}  \ ;
\eean

\item (2) One of the $n_i$ is odd. We will have three sub-cases.
Let us focus on the case $n_1=2k_1, n_2=2k_2, n_3=2k_3+1$. It is
easy to see that $\tr(\sigma_3)=0$. Thus, this category does not give
non-zero meson operators.

\item (3) One of $n_i$ is even. Again there are three sub-cases and we
focus on the case where $n_1=2k_1, n_2=2k_2+1, n_3=2k_3+1$. It is
easy to see that $\tr(\sigma_2 \sigma_3)=0$. Thus the contribution in
this category is again zero.

\item (4) The last case is that all $n_i$ are odd, $n_1=2k_1+1,
  n_2=2k_2+1, n_3=2k_3+1$
Using $\tr(\sigma_1\sigma_2 \sigma_3)\sim \tr(I)\neq 0$, we have
the counting
\bean \sum_{k_i=0}^{\infty} t_1 t_2 t_3
(t_1^2)^{k_1}(t_2^2)^{k_2}(t_3^2)^{k_3}={t_1 t_2 t_3\over (1-t_1^2)
(1-t_2^2)
 (1-t_3^2)} \ .
\eean

\item (5) It can be shown that
$\tr(X^{2k} Y^{2k} Z^{2k})=\tr( (XYZ)^{2k})$, thus we need to be
careful about double-counting. However, categories (1) and (4)
have different powers (even or odd), so we do not have a 
double-counting problem here.

\end{itemize}

Adding the above two together we have the final counting to be,
\[
f_\infty(t_1,t_2,t_3;~\IC^3/\IZ_2^2)_{\mbox{torsion}} = 
{1+t_1
t_2 t_3\over (1-t_1^2) (1-t_2^2)
(1-t_3^2)} \ . 
\]
We should take
the plethystic logarithm to check the equation for moduli space. 
It should be the form
$x y z= t^2$. To see this let us first set $t_1=t_2=t_3=t$ and take
pletytistic logarithm and indeed we get (terminating) polynomial
expression, $3t^2+ t^3-t^6$,
which is exactly what we should have, as one could see from case $m=2$
of \eref{f1-zmzm}. Indeed,
for $N=1$, it does not give a three-dimension moduli space, but,
rather, a degenerate one-dimensional one which is what was argued above
from \cite{Douglas:1999hq}, viz., ${3\over (1-t)^2}$.
We can be more refined and actually compute the full plethystic
logarithm with all three variables, giving us
$t_1^2+t_2^2+t_3^2+t_1 t_2  t_3 - (t_1t_2t_3)^2$.

\subsection{The General $\IZ_n \times \IZ_n$ Case}
Let us proceed to the general case. For the group
$\IZ_n\times \IZ_n$, with action
\bea  
g_1: ~~(z_1, z_2,z_3) & \to & (z_1, e^{-{2\pi i\over n}} z_2,
e^{{2\pi i\over n}} z_3), \\ g_2: ~~(z_1, z_2,z_3) & \to & (
e^{{2\pi i\over n}} z_1, z_2, e^{-{2\pi i\over n}} z_3) \ ,
\eea
the discrete torsion is $\IZ_n$, with the 2-cocycle class
given by $\W\epsilon^m ((a,b),(a',b'))=
\zeta^{m(ab'-a'b)}$ and $\zeta := e^{\pi i\over n}$ for $n$ even or
$\zeta= e^{2\pi i\over n}$ for $n$ odd. We will consider the case that
$gcd(m,n)=1$, for which the projective representation is given
by
\bea 
\gamma_1(g_1)= P,~~~~~\gamma_1(g_2)= Q \ ,
\eea
with $P$ and $Q$ being
the following $n\times n$ matrix (where $\epsilon= \zeta^{2m}$ and
$\epsilon^n=1$)
\bean 
P=\left( \begin{array}{ccccc} 0 & 1 & 0 & \cdots & 0 \\
0 & 0 & 1 & \cdots & 0 \\ \cdots & \cdots & \cdots & \cdots & \cdots
\\ 0 & 0 & \cdots & 0 & 1 \\ 1 & 0 & 0 & \cdots & 0
\end{array} \right),~~~~Q=\left(  \begin{array}{ccccc} 0 & \epsilon &
  0 & \cdots & 0 \\ 
0 & 0 & \epsilon^2 & \cdots & 0 \\ \cdots & \cdots & \cdots & \cdots
& \cdots
\\ 0 & 0 & \cdots & 0 & \epsilon^{n-1} \\ 1 & 0 & 0 & \cdots & 0
\end{array}\right)
\eean
for $n$ odd. For $n$ even, $P$ is the same, while $Q$ is
(with $\delta^2=\epsilon$):
\bean Q=\left(  \begin{array}{ccccc} 0 & \delta & 0 & \cdots & 0 \\
0 & 0 & \delta^{3} & \cdots & 0 \\ \cdots & \cdots & \cdots & \cdots
& \cdots
\\ 0 & 0 & \cdots & 0 & \delta^{2n-3} \\ \delta^{2n-1} & 0 & 0 & \cdots & 0
\end{array}\right) \ .
\eean
We have the following properties
\bea 
PQ= \epsilon Q P,~~~P^n=1=Q^n,~~~\tr(P^k)=\tr(Q^k)=\tr(Q^r
Q^{k-r})=0,~~~if~~k\neq n Z \ .
\eea

Under the condition $gcd(m,n)=1$, the theory has gauge group
$U(M)$, with three chiral adjoint fields $\phi_i$ and superpotential
$\tr(\phi_1 \phi_2 \phi_3-\epsilon^{-1} \phi_1 \phi_3 \phi_2)$. This
gives F-term condition
\bea 
\phi_i \phi_j-\epsilon^{-1}\phi_j
\phi_i,~~~(i,j)=(1,2),(2,3),(3,1) \ .
\eea
Again, the solution of F-terms and D-terms relation is given by
\bea 
\phi_1= X\otimes Q,~~~\phi_2=Y\otimes P,~~~~\phi_3=Z\otimes
(QP)^{-1} \ ,
\eea
where $X,Y,Z$ all commute, just like parent $\cN=4$ theory. Now, we
consider the mesonic operators
$\tr(\phi_1^{n_1} \phi_2^{n_2} \phi_3^{n_3})$ and
write $n_i= n k_i+ s_i$ with $k_i=0,...,\infty$ and $s_i=0,...,n-1$.
The key part is to see if $\tr(Q^{s_1} P^{s_2} (QP)^{-s_3})$ is zero
(where we have used the fact that $P^n=Q^n=\II$).

To see the properties of $P,Q$ we use the following observation. We take
$P,Q$ as the action of $n$-dimensional vector space with basis
$e_0,...,e_{n-1}$. Then the action is as follows (for simplicity we
assume $n$ is odd)
\bea   
P(e_i)= e_{i+1},~~~~Q(e_i)= \epsilon^{i+1} e_{i+1} \ ,
\eea
whence we have
\bean 
P^k(e_i) &= & e_{i+k},~~~~\tr(P^k)=0,~~if~~k \neq n Z \\
Q^{k} (e_i) & = & (\prod_{r=1}^{k} \epsilon^{i+r}) (e_{i+k})=
\epsilon^{ki + {k(k+1)\over 2}}(e_{i+k}), ~~\tr(Q^k)=0,~~if~~k \neq n
Z  \\
P^{r_1} Q^{r_2} (e_i) & = & P^{r_1} \epsilon^{r_2 i +
{r_2(r_2+1)\over 2}}(e_{i+r_2})= \epsilon^{r_2 i + {r_2(r_2+1)\over
2}} (e_{i+r_1+r_2}), \\
\tr(P^{r_1} Q^{r_2}) & = & \delta(r_1+r_2- n Z) \sum_{i=0}^{n-1}
\epsilon^{r_2 i + {r_2(r_2+1)\over 2}}= \epsilon^{r_2 i +
{r_2(r_2+1)\over 2}}\delta(r_1+r_2- n Z) \delta(r_2-n \W Z)\\
& = &\epsilon^{r_2 i + {r_2(r_2+1)\over 2}}\delta(r_1- n Z_1)
\delta(r_2-n  Z_2) \ .
\eean
The last equation is very important; it tells us that we need
$r_1,r_2$ to be integer when multiplying by $n$.

Now, we calculate
\bean (QP)^r & = & \epsilon^{ r(r-1)\over 2} Q^r P^r\\Q^{s_1}
P^{s_2} (QP)^{-s_3} & =
& Q^{s_1} P^{s_2} \epsilon^{{s_3(s_3+1)\over 2}} Q^{-s_3} P^{-s_3}  \\
& \sim &  P^{s_2-s_3} Q^{s_1-s_3} \ .
\eean
From this we can see that $\tr(Q^{s_1} P^{s_2} (QP)^{-s_3})\neq 0$
when and only when $s_1-s_3=n Z_1, s_2- s_3= n Z_2$. Also, because
$s_i\in [0,n-1]$ we have the only possibility that $Z_1=Z_2=0$.

With all these analysis we have following counting
\bean 
f_\infty(t_1,t_2,t_3;~\IC^3/\IZ_n^2)_{\mbox{torsion}}
&= & \sum_{k_i=0}^{\infty} \sum_{s_i=0}^{n-1} \sum_{Z_1,Z_2}
t_1^{n k_1+s_1} t_2^{n k_2+s_2} t_3^{n k_3+s_3}
\delta( s_1-s_3-n Z_1)\delta(s_2-s_3-n Z_2 )\\
& = & {1\over (1-t_1^n)(1-t_2^n)(1-t_3^n)} {(1- (t_1 t_2
t_3)^n)\over (1-t_1 t_2 t_3)} \ .
\eean
First, for simplicity we can set all $t_i$ to be equal to $t$ and get
${ (1-t^{3n})\over (1-t^3)(1-t^n)^3}$.
For $n=2$ it goes back to our previous result. Taking the
plethytistic logarithm we get the (terminated) polynomial $t^3+ 3
t^n-t^{3n}$ which is the equation of $ xy z= t^n$ with $t$ degree
three and $x,y,z$ degree $n$. We can also compute the full plethystic
logarithm including the three variables and obtain
$t_1^n+t_2^n+t_3^n+t_1t_2t_3-(t_1t_2t_3)^n$.

It is very interesting to notice that, comparing with \eref{zmzmf} and
 \eref{zmzmf1}, the result for $\IZ_n\times \IZ_n$ with torsion
is same as the one without discrete torsion. This is consistent with
our claim that at least for complete intersection geometries we
can get the fundamental invariant 
$g_1(t)$ using defining equation directly. And indeed, the
defining equation does not distinguish if there is torsion or not.

In fact, for an orbifold action, there are three parts:
(1) The adjoint action on Chan-Paton factors; (2) The space-time
action on the three chiral multiples $X,Y,Z$; and (3) The projected
superpotential coming from Clebesh-Gordon coefficients.
To get the quiver, we need to know the information of the
first two parts only.
The space-time action is always a faithful representation while the
Chan-Paton action could be projective. However, because it is an
adjoint action, the cocycle factor does not affect the discussion
and this is why we can use the covering group to get the quiver diagram.
The difference between faithful representation and projective
representation is that the dimension of matrix is different, thus,
given $N$ D3-brane probes we have less number of gauge groups under
projective representation. For example, for $\IZ_n\times \IZ_n$, in the
minimum case we have only one gauge group for projective
representation while for the faithful one (without discrete torsion), 
we have $U(1)^{n^2}$ gauge groups.

To determine the theory completely, we need to know the superpotential
as well; now, the
difference between projective and faithful
representations gives different F-term relations. 
However, as we have emphasized that the space-time action is the same with or 
without discrete torsion, it is reasonable that we get the same answer and
the Molien (Hilbert) series seems to apply. 
Although we have not checked all cases 
where the geometry may or may not be complete intersection, we do conjecture 
that counting will be same in all such cases.

%
%
\section{Hilbert Schemes and Symmetric Products}\label{s:hilb}\setequation
We have delved into orbifolds and quotients quite intensively in the
foregoing discussions. Of key significance in our derivation in
\cite{Benvenuti:2006qr} for
$g(\nu,t)$ is that the full generating function for all finite $N$
relies on an important quotient, viz., the space
\beq
\sym^N(\cM) := \cM^N/S_N \ , 
\eeq
the $N$-th
symmetric product of a space $\cM$. This is important because if $\cM$ is
the vacuum moduli space of a single D3-brane probe (i.e., the
transverse Calabi-Yau singularity), then $\cM^N/S_N$ is
that of a stack thereof.
In general, $\cM^N/S_N$ may have singularities, however, there is a
canonical resolution called the Hilbert scheme \cite{nakajima}, 
which is much richer
in structure than the mere moduli space of $N$ (non-coinciding) 
points in $\cM$ as captured by $\cM^N/S_N$. 

Formally, the Hilbert scheme is defined to be the set of all
sub-schemes (we switch liberally between schemes and ideals using the
algebra-geometry correspondence) of a variety $X$ of length $N$, i.e.,
\beq
\hilb^n(X) := \left\{ \mbox{ideals }I \subset X | \dim(X/I) = N
\right\} \ .
\eeq
In dimension 1, everything is easy and we have that
\beq
\hilb^N(\IC) = \sym^N(\IC), \qquad
\hilb^N(\IP^1) = \IP^N \ .
\eeq

\subsection{The Second Symmetric Product of $\IC^m$}
Let us see what we can say about $\IC^m$. We know that
\beq\label{gN-symg1}
g_N(t;~\cM) = g_1(t;~\sym^N(\cM)),
\eeq
and that
\beq
g_1(t;~\IC^m) = f_\infty(t;~\IC^m) = \frac{1}{(1-t)^m} \ .
\eeq
In fact, the expression is refined as
$\prod\limits_{i=1}^m \frac{1}{(1-t_i)}$.
Some immediate results can be read off for $N=2$. Here, $S_2 \simeq
\IZ_2$. Take $m=2$, we have that
\beq
\sym^2(\IC^2) \simeq \IC[x_1,y_1,x_2,y_2]^{\IZ_2}, \qquad
\IZ_2 \simeq \gen{\tmat{
   0 & 0 & 1 & 0 \cr 0 & 0 & 0 & 1 \cr 1 & 0 & 0 & 
   0 \cr 0 & 1 & 0 & 0 \cr  }} \ ;
\eeq
so the Molien series and the corresponding plethystic logarithm are simply
\beq\label{sym2c2}
M(t) = f_\infty(t) = \frac{1 + t^2}
  {{\left( 1 - t \right) }^4\,
    {\left( 1 + t \right) }^2} \ , \qquad
f_1(t) = PE^{-1}[f(t)] = 2\,t + 3\,t^2 - t^4 \ .
\eeq
Again, this has a refinement and $f_1 = t_1 + t_2 + t_1^2 + t_1 t_2 + t_2^2
- t_1^2 t_2^2$.
This is consistent with the fact that the defining equation for $\IC^4
/ \IZ_2$ with our chosen action, is a complete intersection. It is
given as a single relation (of degree 4) amongst 5 primitive (2
linear and 3 quadratic) invariants (the syzygies can be readily
computed using \cite{m2}):
\bea
Y_{1,\ldots,5} &:=& \{ x_1 + x_2,
   y_1 + y_2,
   {x_1}^2 + {x_2}^2,
   x_1\,y_1 + x_2\,y_2,
  {y_1}^2 + {y_2}^2 \}; \\ \nn
&\Rightarrow& {Y_2}^2\,Y_3 - 2\,Y_1\,Y_2\,Y_4 + 2\,{Y_4}^2 + 
   {Y_1}^2\,Y_5 - 2\,Y_3\,Y_5 = 0 \ .
\eea
It is clear here that the refinement comes from an independent
counting of $x$'s and $y$'s.
Set $t_1$ to count $x$'s, $t_2$ to count $y$'s and the equation for
$f_1$ follows.

Next, we can take $m=3$ and $N=2$. Now, we have that
\beq
\sym^2(\IC^3) \simeq \IC[x_1,y_1,z_1,x_2,y_2,z_2]^{\IZ_2}, \qquad
\IZ_2 \simeq \gen{\tmat{
   0 & 0 & 0 & 1 & 0 & 0 \cr 0 & 0 & 0 & 0 & 1 & 0 \cr 
   0 & 0 & 0 & 0 & 0 & 1 \cr 1 & 0 & 0 & 0 & 0 & 0 \cr 
   0 & 1 & 0 & 0 & 0 & 0 \cr 0 & 0 & 1 & 0 & 0 & 0 \cr
    }} \ ,
\eeq
giving us
\beq
M(t) = f_\infty(t) = \frac{1 + 3\,t^2}
  {{\left( 1 - t \right) }^6\,
    {\left( 1 + t \right) }^3} \ ;
\qquad
f_1(t) =
3\,t + 6\,t^2 - 6\,t^4 + 8\,t^6 - 18\,t^8 + 
  \cO(t^{10})
\ .
\eeq
Already, here, we see from the expression for $f_1$ 
that the space is not a complete intersection.
Here, the refinement is interesting:
\beq
g_1 = \frac{1+t_1 t_2 + t_2 t_3 + t_3 t_1}{ \prod_{i=1}^3 (1-t_i)
  (1-t_i^2)} \ ,
\eeq
giving
\beq
t_1+t_2+t_3 + \sum\limits_{i \le j} 
t_i t_j - \sum\limits_{i<j} (t_i t_j)^2 - t_1 t_2 t_3 (t_1+t_2+t_3) + ...
\eeq

A general formula for $g_2$ of any $m$ is the following:
\[
g_2({t_1,\ldots,t_m} ; \IC^m) = 
\left(\prod_{i=1}^m (1-t_i) (1-t_i^2)\right)^{-1}
\left( 1 + \sum\limits_{i<j} t_i t_j + \sum\limits_{i<j<k<l} 
t_i t_j t_k t_l + 
\ba{c}\mbox{Similar order}\\ \mbox{6 term}\ea + \ldots \right) 
\]
As a historical digression, the series coefficients of $M(t)$ for
$\sym^2$ of $\IC^m$ are known
as Paraffin (or Alkane) Numbers, having to do with reading off diagonals of
Losanitsch's generalisation of Pascal's triangle
\cite{sloane,Losanitsch}.

\subsection{The $n$-th Symmetric Product of $\IC^2$}
Alternatively, one could analyse the family
\[
\sym^n(\IC^2) \simeq \IC[x_1,y_1; \ x_2,y_2; \ \ldots; \ x_n,y_n] /
S_n \ ,
\]
where $(x,y)$ are the coordinates of $\IC^2$ and $S_n$ permutes the
$n$-tuple of points $(x_i,y_i)$. This is very much in the spirit of
Hilbert scheme of points on surfaces as detailed in \cite{nakajima}.
In fact, in this case, \cite{naka-yoshi} has given, in our
notation, the $\nu$-inserted plethystic exponential (cf.~Eq (4.5),
{\it cit. Ibid.}):
\beq
g(\nu,~t_1,t_2;~\IC^2) = PE_\nu[((1-t_1)(1-t_2))^{-1}] = \exp\left[
\sum_{k=1}^\infty \frac{\nu^k}{k(1-t_1^k)(1-t_2^k)} 
\right] \ .
\eeq
Unrefining by setting $t_1 = t_2 = t$ and power expanding in $\nu$,
gives, for the coefficient of $\nu^n$, the Hilbert series for
$\sym^n(\IC^2)$:
\beq\label{symN-C2}
 PE_\nu[\frac{1}{(1-t)^2}] = 1 + \frac{1}{(1-t)^2} \nu +
\frac{1 + t^2}{(1-t)^4(1+t)^2} \nu^2 +
\frac{1 + t^2 + 2\,t^3 + t^4 + t^6}
  {{\left( 1 - t \right) }^4\,
    {\left( 1 + t \right) }^2\,
    {\left( 1 - t^3 \right) }^2} \nu^3 + \cO(\nu^4)
\eeq
We see that the $\nu$-term is indeed that of $\sym^1(\IC^2) = \IC^2$
and the $\nu^2$-term is what we calculated in \eref{sym2c2} for
$\sym^2(\IC^2)$. 

This is, of course, in perfect congruence with our proposal in
\cite{Benvenuti:2006qr}: that the $\nu$-inserted plethystic
exponential of the Hilbert series should give the generating
function for $g_N$, the multi-trace generating function for $N$
D3-branes. Indeed, the coefficient to $\nu^N$ in \eref{symN-C2}
is the Hilbert series $f_\infty = g_1$ of $\sym^N(\IC^2)$. 
However, recalling from \eref{gN-symg1} that $g_1(t;~\sym^N(\IC^2)) 
= g_N(t;~\IC^2)$, the agreement is re-assuring.

In fact, \cite{sloane}, the series-expansion in $t$ for
the $n$-th coeffcient, i.e., the $t$-expansion for the Hilbert
series of $\sym^n(\IC^2)$ gives the planar integer partitions
(Young-Tableaux) of trace $n$. That is to say, the coefficient of $t^m$
corresponds to the number of ways of writing
the given integer $m$ as $\sum_{i,j} z_{i,j}$ with $z_{i,j} \in \IZ_+$
such that $z_{i+1,j} > z_{i,j}$,  $z_{i,j+1} > z_{i,j}$ and 
$\sum_{i,i} z_{i,i} = n$.

\newpage

%
%
%
\section{A Detailed Analysis of $Y^{p,q}$}
\begin{figure}
\centerline{\epsfxsize=4in\epsfbox{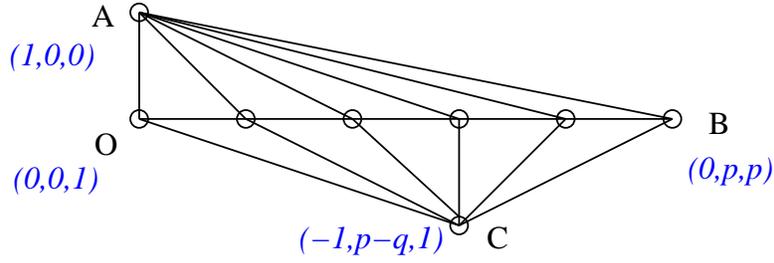}} 
\caption{{\sf The toric diagram for the space $Y^{p,q}$.}}
\label{f:Ypq}
\end{figure}

Having indulged ourselves with quotient spaces, let us change our
palatte awhile to toric varieties.
The space $Y^{p,q}$ 
(cf.~e.g.~\cite{Gauntlett:2004zh,Gauntlett:2004yd,Martelli:2004wu,Benvenuti:2004dy,Benvenuti:2004wx})
was studied in detail in \cite{Benvenuti:2006qr}.
The toric data is given by the integer lattice points
$O=(0,0,1)$, $A=(1,0,1)$, $B=(0,p,1)$ and $C=(-1,p-q,1)$ as drawn in
\fref{f:Ypq}.
As indicated, we take the trianglization of the toric diagram by
connecting the point
$T_a:=(0,a,1)$ to $A$ and $C$ with $a=1,...,p$ (so that $T_p=B$). 
Thus, we have $2p$ triangles given by $T_a A T_{a+1}$ and $T_a C
T_{a+1}$, $a=0,...,p-1$. From this we can read out the fundamental 
generating function $f_\infty$ (cf.~\cite{Benvenuti:2006qr}) as
\beq\label{p-ypq} \hspace{-1cm} 
f(x,y,z;~Y^{p,q}) = \sum_{a=0}^{p-1}
\frac{1}{\left( 1 - x \right) \,
    \left( 1 - \frac{x^a\,y}{z^a} \right) \,
    \left( 1 - \frac{x^{-1 - a}\,z^{1 + a}}{y} \right) } +
\frac{1}{\left( 1 - \frac{1}{x} \right) \,
    \left( 1 - \frac{x^{-a + p - q}\,y}{z^a} \right) \,
    \left( 1 - \frac{x^{1 + a - p + q}\,z^{1 + a}}
       {y} \right) } \ .
\eeq
To get a more compact expression for $f$, we need to
sum up the two series. This can be done as follows. First, for each
term in the series $OAB$ we can write it as
\bean 
{-x z^{-1}\over (1-x)(1-x z^{-1})}
\left( -{1\over 1-x^{-a} z^{a} y^{-1}}+{1\over 1-x^{-a-1} z^{a+1}
  y^{-1}}\right) \ .
\eean
Summing up from $a=0$ to $a=p-1$ we get
\[
{x y(z^p-x^p)\over (1-x)(z-x)(1-y) (x^p y-z^p)} \ .
\]
For the series $OBC$, each term can be written as
\bean 
{-x\over (1-x) (1-x z)}\left( {1\over 1-x^{-a+p-q} y z^{-a}}
-{1\over 1-x^{-a-1+p-q} y z^{-a-1}}\right) \ ,
\eean
and summing up we get
\[ 
{x y(1-x^p z^p)\over (1-x) (1-x z) (1- x^{p-q} y) (x^q z^p-y)} \ .
\]
Putting all together we have
\beq
f(x,y,z;~Y^{p,q}) =  {x y\over (1-x)}\left( {(1-x^p z^p)\over
(1-x z) (1- x^{p-q} y) (x^q z^p-y)}+{(z^p-x^p)\over (z-x)(1-y) (x^p
y-z^p)}\right) \ .
\label{Ypq-gen} 
\eeq

Now, let us try to find the defining equation for $Y^{p,q}$ using the
plethystic programme. The basic
invariants and relations were counted in \cite{Pinansky:2005ex} and 
we wish to use our plethystic programme to check those results. We
therefore need the generating function $f_1$ by 
taking the plethystic logarithm of \eref{Ypq-gen}.

For this purpose let us first study the structure of the dual cone,
which in toric geometry will give us the relations. It is
easy to find that
in our coordinates the following dual vectors are generators of dual cone:
\[ 
e_1=(0,1,0),~~~~e_2=(-p,-1,p),~~~~e_3=(q,-1,p),~~~~~e_4=(p-q,1,0) \ .
\]
However, these are the generators of the dual cone over $\IR_+$,
we need the generators over $\IZ_+$.
This is to say that the above four vectors are not complete in the
sense of lattice 
points, and there will exist some integer vectors, which are linear 
combinations of
these four with positive real number coefficients, but not with
positive integer coefficients.

To find these missing vectors, first we notice  
that $e_3-e_2= (p+q,0,0)$, thus the following $(p+q+1)$ vectors
\[ 
e_{23,m} =e_2+ (m,0,0)= (-p+m,-1,p),~~~~m=0,...,p+q
\]
must be included into the $\IZ_+$-generators of the dual cone.
Similarly, since $e_4-e_1=(p-q,0,0)$, the following $(p-q+1)$ vectors
should also be included:
\[  
e_{14,m}=(m,1,0),~~~~~m=0,...,p-q \ .
\]
We are not finished yet. From the fact that
$e_3+e_4=(p,0,p)$ we find that we
need $(1,0,1)$ and from $e_1+e_2=(-p,0,p)$, we need
$(-1,0,1)$. Finally,
from these two we get $(0,0,2)$ so we need $(0,0,1)$ as generator.
Thus we have three more generators
\[ 
e_{5}=(1,0,1),~~~e_6=(-1,0,1),~~~~~e_7=(0,0,1) \ .
\]
Putting all together  we have $2p+5$ generators $e_{23,m}, e_{14,m},
e_{5}, e_6, e_7$ as claimed by \cite{Pinansky:2005ex}.

To have a simple result for the plethystic logarithm, we want as many as
generators having same degree. One simple choice could be the following
scaling: $x\to 1, y \to t^p, z\to t^2$ under which the
generators $e_{23,m}$ and $e_{14,m}$ have same degree. However, when
$p$ is even, there is an interference
between the number of variables and
the number of equations for the definition of geometry. By this we
mean that in expanding the expression for the plethystic logarithm,
positive terms signify invariants while negative terms signify
relations, these could potentially cancel and confuse the counting;
this situation was encountered in the example of the non-complete
intersection $\IC^3/\IZ_3$ in \cite{Benvenuti:2006qr}.
Thus the above scaling is
not a good choice. Another choice will be the $x\to 1, y \to
t^{2p-1}, z\to t^4$. In this case, $e_{23,m}$ and $e_{14,m}$ do not
have same degree, but the interference is avoided.

However, we can do better by having multiple variables in the
plethystic logarithm. To do so, first we need to set $x\to 1$.  The
reason is very simple because we want to set $e_{23,m}$ (as well as
$e_{14,m}$) to have the same degree. After this we have
\bean
f(x\to 1,y,z;~Y^{p,q}) &=& { y A+ y^2 B+ y^3 C\over (1-y)^2(1-z)^2
 (y-z^p)^2} \ ; \\
A &:=&  -z^p ( z^{p+1} (p-q-1)-z^p (p-q+1)+z(p+q+1)-(p+q-1)) \\
B &:=&  (1+z)(1-z^{2p})-4 p z^p(1-z) \\ 
C &:=&  -z^{p+1} (p+q-1)+z^p(p+q+1)-z(p-q+1)+(p-q-1) \ .
\eean
In this form, it is not suitable to take the plethystic logarithm
because of the overall factor $y$ in the numerator as well as the factor
$(y-z^p)$ in the denominator, which would give a logarithmic singularity
in at $y=0$. To amend this, we change variables as $y\to y$ and $z\to
t y$. Thus
\bean 
&&f(x\to 1,y,t y;~Y^{p,q}) = {  A+  B+ y C\over (1-y)^2(1-t y)^2
  (1-t^p y^{p-1})^2} \\ 
&&A := - t^p y^{p-1} ( ( t y)^{p+1} (p-q-1)-( t y)^p (p-q+1)+ t
  y(p+q+1)-(p+q-1)) \\
&&B := (1+t y)(1-( t y)^{2p})-4 p ( t y)^p(1- t y) \\
&&C := -( t y)^{p+1} (p+q-1)+( t y)^p (p+q+1)- t y(p-q+1)+(p-q-1) \ .
\eean
Now, we can take the plethystic logarithm and get the right answer! 
It is easy to
see that under the above scaling we have $(p+q+1)$ variables with
scaling $e_{23,m}\to y^{p-1} t^p$,
$(p-q+1)$ variables with scaling $e_{14,m}\to y$ and three variables
with scaling $e_{5,6,7}\to y t$.

Let us check the above result with the tabulation of the
following several examples:
\[
\ba{c|c}
(p,q) & f_1(t,y; Y^{p,q}) \\ \hline
(p=1,q=0) & ( 2t+2 y)- ty \\
(p=1,q=1) & (3 t + y) - t^2 \\
(p=2,q=0) & (3 t^2 y + 3 y +3 t y) -y^2- 4 t y^2-10 t^2 y^2 -4 t^3
y^2-t^4 y^2 +...\\ 
(p=2,q=1) & (4 t^2 y+ 2y+ 3 t y)-2 t y^2-9 t^2 y^2-6 t^3 y^2-3 t^4 y^2+... \\
(p=2,q=2) & (5 t^2 y+ y+3 t y)- 6 t^2 y^2-8 t^3 y^2-6 t^4 y^2+... \\
(p=3,q=0) & (4 t^3 y^2+ 4 y+ 3 t y)- 3 y^2-6 t y^2-t ^2 y^2-16 t^3
y^3-6 t^4 y^3-3 t^6 y^4 +... \\ 
(p=3,q=1) & (5 t^3 y^2+3 y +3 t y)-y^2-4 t y^2-t^2 y^2-15 t^3 y^3-8
t^4 y^3-6 t^6 y^4 +... \\ 
(p=3,q=2) & (6 t^3 y^2+2 y+ 3 t y)- 2 t y^2 -t^2 y^2-12 t^3 y^3-10 t^4
y^3-10 t^6 y^4+...\\ 
(p=3,q=3) & (7 t^3 y^2+y+3 t y)-t^2 y^2 -7 t^3 y^3-12 t^4 y^3-15 t^6
y^4+...
\ea
\]
The interpretation of $f_1$ was outlined in \cite{Benvenuti:2006qr}.
For $p=2, q=0$, for example, we should
have $3$ variables with scaling $t^2 y$, $3$ variables with scaling
$y$ and another $3$ with scaling $t y$; this is 
given exactly inside the first bracket. 
The remaining part is
the information about relations, i.e, defining equations. 
The term $-y^2$ means
there is one relation among $3$ variables with scaling $y$.
Similarly the term $-4 t y^2$ tells us that there are  four
relations between variables with scaling $y$ and variables with
scaling $yt$.

Checking all examples we find that the total
number of defining equations is given by $p^2+10p-4$. It is
different from the claim given by  \cite{Pinansky:2005ex}
where it is claimed that the
number of equations should be $p^2+10p-q^2-4$. 
However, in {\it cit.~ibid.}, only minimum relations are counted. 
In other words, if $a=b$ and $b=c$ then $a=c$
will not be counted as a new relation. However, in our case, $a=c$
would be counted separately.

\newpage

\section{Conclusions and Prospects}
We have proposed a {\bf plethystic programme} for the counting of
gauge invariant operators in gauge theories. Though we have
restricted our attention to chiral BPS mesonic operators in
world-volume theories of D-branes probing Calabi-Yau singularities,
the programme should be of wider applicability. In the case of our
present focus, an intimate web of connexions between geometry, gauge
theory and combinatorics emerges. This field of quiver theories is
where the plethystics fully blossom. In a way, this does not
surprise us. Indeed, for D-brane quiver theories, the mesonic gauge
invariants in the chiral ring, modulo the F-term constraints, should
give a classical moduli space that by construction is the Calabi-Yau
variety which the brane probes.

What is beautiful about the plethystic programme is that the
plethystic exponential function and its inverse provide the explicit
link between the geometry and the gauge invariants. One only needs
to construct a fundamental invariant of the Calabi-Yau manifold
$\cM$, which we have called $f = f_\infty = g_1$ and which
mathematically corresponds to the Hilbert series. This is the
generating function for the single-trace operators. The plethystic
logarithm, $PE^{-1}$, gives all the syzygies of $\cM$. In the case
of $\cM$ being complete intersection, $f_1 = PE^{-1}[f]$ is a
polynomial from which one immediately reads out the defining
equation of $\cM$. On the other hand, the plethystic exponential
gives $g_\infty = PE[f]$, the generating function for the
multi-trace operators.

Continuing with \cite{Benvenuti:2006qr}, we have provided a host of
examples to demonstrate the power of the plethystic programme,
ranging from orbifold theories to toric singularities, from discrete
torsion to Hilbert schemes, touching upon such interesting curios as
relations to Young Tableaux and to the MacMahon Conjecture.
Importantly, we have also, using results of Temperley and
Haselgrove, generalised the formulae of Hardy-Ramanujan and
Meinardus, in estimating the asymptotic behaviour of the number of
such operators. This is an estimate of the degrees of freedom of the
gauge theory, whereby providing us with explicit expressions for the
entropy for an arbitrary number $N$ of D-branes.

We have, of course, only touched upon the fringe of a fertile
ground. How do plethystics teach us about other branches of the
moduli space? For example, it will be interesting to count baryonic
branch and to see if syzygies of  divisors wrapped by D3-branes be
captured by plethystics.
Including of baryonic operators into the counting should correspond
to the geometry of cycles in the Sasaki-Einstein manifold, can the
syzygies of these divisors be captured by plethystics? 
Recently, progress in counting baryonic
operators has been made in \cite{Butti:2006au}. Along similar lines is
the mixed branch studied in \cite{HR}. Can our programme be extended
to study such other branches? What about 1/4 or 1/8 BPS states? 

Moreover, in \cite{Benvenuti:2006qr}, we have performed
the counting for geometries whose quiver theories have not yet been
constructed; we could also do so for non-quiver and even non-SUSY
theories. 
Indeed, there has also been much study of certain indices
of superconformal theories
\cite{Kinney:2005ej,Romelsberger:2005eg,Biswas:2006tj,Mandal:2006tk,
Sinha:2006ac,Nakayama:2005mf} as well as the counting of
instantons such as in \cite{Fucito:2006kn}. How does our plethystic
programme relate to these counting problems?
The portals to a Grecian mansion have been opened to us,
to fully explore the plethora of her plethystic secrets shall be our
continued goal.


\section*{Acknowledgements}
We heartily acknowledge Nemani Suryanarayana and Balazs Szendroi
for wonderful communications.
B.~F.~is obliged to the Marie Curie Research
Training Network under contract number MRTN-CT-2004-005104 and once
more extends
his thanks to Merton College, Oxford for warm reception at the final
stages of the draft. 
A.~H.~is grateful to Sergio Benvenuti, Kazutoshi Ohta, Toshio Nakatsu,
Yui Noma, and Christian Romelsberger for
enlightening discussions.
Y.~H.~H. kisses the grounds of Merton College, Oxford,
to whose gracious patronage, through the FitzJames Fellowship, he owes
his inspiration, and he kisses the hands of Senyorita
L.~Figuerola-Sol\'e, whose eyes, like rays of Mediterranean sun,
pierces the scholastic melancholy of his soul.

\newpage


\end{document}